\documentclass[nofootinbib,prd,aps,nosuperscriptaddress]{revtex4-2}

\usepackage[utf8]{inputenc}
\usepackage[T1]{fontenc}
\usepackage{lmodern}
\usepackage{graphicx}
\usepackage{enumerate}
\usepackage{titlesec, blindtext, color}
\usepackage{hyperref}
\usepackage{amsmath}
\usepackage{amssymb}
\usepackage{amsthm}
\usepackage{bbm}
\usepackage {tensor}
\usepackage{multirow}
\usepackage{etoolbox}
\usepackage{enumitem}
\usepackage[english]{babel}
\usepackage{dsfont}
\usepackage{float}

\newcommand{\beq}{\begin{equation}}
\newcommand{\eeq}{\end{equation}}
\newcommand{\beqa}{\begin{eqnarray}}
\newcommand{\eeqa}{\end{eqnarray}}

\newcommand{\mz}[1]{{#1}}

\begin{document}

\title{Gravastar-like black hole solutions in $q$-theory}
\author{M. Selch}
\affiliation{Ariel University, Ariel, 40700, Israel}

\author{J. Miller}
\affiliation{Ariel University, Ariel, 40700, Israel}

\author{ M.A.Zubkov }
\affiliation{Ariel University, Ariel, 40700, Israel}

\date{\today}
\begin{abstract}\noindent 
We present a stationary spherically symmetric solution  of the Einstein equations, with a source generated by a scalar field 
of  
$q$-theory. 
In this theory Riemannian gravity, as described by the Einstein - Hilbert action, is coupled to a three - form field that
describes the dynamical vacuum. Formally it behaves like a  matter field with its own stress - energy tensor, equivalent to a scalar field minimally coupled to gravity. 
The asymptotically flat solutions obtained to the field equations 
represent black holes. For a sufficiently large horizon radius the energy density is localized within a thin spherical shell situated just outside of the horizon,
analogous to a gravastar.
The resulting solutions to the field equations, which   admit this class of 
configurations, satisfy existence conditions that stem from the  Black Hole no - hair theorem,
thanks to the presence of a region in space in which the energy density is negative. 
\end{abstract}
\maketitle
\section{Introduction}\noindent
General Relativity (GR)
is  
based on the axiom
that the gravitational field is encoded by the geometry of spacetime:
in a perfect vacuum spacetime is flat, while 
massive objects distort the surrounding spacetime from being otherwise flat to having a curved geometry.
Test particles move  along  geodesics of the background spacetime in a way such that
the trajectory depends on the geometry of the spacetime. In flat space the trajectory is a straight line, while
in curved space it gets  accelerated away from  straight line motion.
This is  the \emph{equivalence principle}: the gravitational field is equivalent to acceleration in the background spacetime.
In this manner Newton's `action at a distance' theory of gravity is replaced by Einstein's field theory approach, in which  the field is the very 
geometry of the spacetime.
\par 
The vacuum Einstein field equations contain a cosmological constant  \cite{Einstein:1917ce}, 
which 
relates the large-scale expansion of the universe in cosmological models.
In turn the cosmological constant is related to the 
vacuum energy density \cite{Bronstein:1936syj,Landau:1933utz,Zeldovich:1967gd,Weinberg:1972kfs,Veltman:1975vx}.
Its value according to  astronomical observations
lies at a typical energy scale of the order of  $10^{-3}\;\text{eV}$
\cite{Boomerang:2000efg,WMAP:2006jqi,Riess:2006fw},
while its range of values as inferred from theoretical models 
is much larger \cite{Weinberg:1988cp,Sahni:1999gb,Padmanabhan:2002ji,Nobbenhuis:2004wn,Polchinski:2006gy}.
\par 
Klinkhamer and Volovik have suggested  
\cite{Klinkhamer:2007pe,Klinkhamer:2008ns} that the 
smallness of the observed vacuum energy density
can be explained on the basis of a thermodynamic argument by which the vacuum energy density is exactly canceled in equilibrium (perfect quantum vacuum). $q$-theory appears as a low-energy effective theory
\cite{Volovik:2007vs,Froggatt:2005jb,Bjorken:2001pe,Giudice:2007qj} as opposed to a purely fundamental theory
\cite{Weinberg:1988cp,Sahni:1999gb,Padmanabhan:2002ji,Nobbenhuis:2004wn,Polchinski:2006gy}, and as a result
contributions to the vacuum energy density become suppressed at macroscopic scales and reside in small perturbations of the equilibrium state.
In the Klinkhamer Volovik model, the 
effective vacuum energy density that enters the low-energy field equations
is described by a vacuum field variable.
At high energies it may be described by the $3$-form field $A_{\beta\gamma\delta}$, which is antisymmetric with respect to permutation of indices.
From this tensor the scalar $q$-field, the low energy vacuum variable, is composed as $q^2=-\frac{1}{24}F_{\alpha\beta\gamma\delta}F^{\alpha\beta\gamma\delta}$, where $F_{\alpha\beta\gamma\delta}=\nabla_{[\alpha}A_{\beta\gamma\delta]}$ is the field strength. 
The equilibrium value of $q$ alters 
if the vacuum is perturbed towards a new equilibrium state. 
More details can be found in \S\ref{sec_model_under_constructn}.
\par 
Black holes (BHs) are unstable due to several effects. They may evaporate gradually resulting in Hawking radiation \cite{Hawking:1975vcx}. 
Alternatively a BH may undergo a transition to a white hole 
\cite{Soltani:2021zmv}, through quantum mechanical tunneling from inside a  trapped region to an  anti-trapped region.
Another possible outcome is 
the formation of a 
\emph{vacuum star}.
In this model the event horizon
takes the form of a boundary between different phases 
of the quantum vacuum \cite{Chapline:2000en}.  
There are a number of similar phenomena between semi-metals and BHs, for example  
 the event horizon emerging on the
boundary between type I and type II Weyl semimetals.
Volovik \cite{Volovik:2021qvc}
has discussed an analogous process that occurs in Dirac and Weyl semimetals
that suggests 
the viability of the formation of 
a \emph{gravastar} or \emph{vacuum star} after vacuum reconstruction, once  Hawking radiation has been ended.
According to \cite{Volovik:2021qvc} (and unlike  conventional gravastars \cite{Mazur:2001fv}) the gravastar admits three distinct regions:
the vacuum inside the Cauchy horizon 
with the  
de Sitter metric, the vacuum inside the thin shell 
between the Cauchy horizon and the event horizon, and
the vacuum outside  the event horizon with the ordinary Schwarzschild metric. The similar objects were also considered in \cite{Khlopov5} (see also \cite{Khlopov1,Khlopov2,Khlopov3,Khlopov4}). Even classically, BHs may be unstable. 
Regarding it as an `excited state',  decay due to exponential growth of the metric (plus soliton) perturbations becomes possible \cite{Ashtekar:2000nx}.
\par 
The geometry of the spacetime in a neighbourhood of a gravastar can be described using 
the action of a $q$-field coupled to gravity, where the $q$-field, as mentioned above,
is related to the energy density of the quantum vacuum \cite{Hawking:1984hk,Aurilia:1980xj}.
The goal of this work is to 
show (by numerical means) that (static and spherically symmetric) ``scalar-haired'' BHs exist within $q$-theory induced by the scalar $q$-field minimally coupled to gravity. The solutions are discussed thoroughly and are interpreted within the context of gravastars.
A similar (numerical) calculation by Nucamendi and Salgado can be found in refs.
\cite{Nucamendi:1995ex}, in which solutions to the field equations are derived 
for the case of a scalar field coupled minimally as well, in a generic static, spherically symmetric and
asymptotically flat spacetime very similar but not identical to our considerations within $q$-theory.
There, ``scalar-hair'' BH solutions  
were shown to exist.
\par 
As mentioned above, 
such BH solutions admit non-trivial
``hair'' associated with the scalar field.
A general BH ``no-hair'' conjecture was originally proposed by Ruffini and Wheeler \cite{Ruffini:1971bza} (see also Hawking 1975 \cite{Hawking:1975vcx} for
a thorough pedagogical overview).
A set of conditions arise from no-hair theorems \cite{Heusler:1996ft}
for the existence of a solution to the Einstein equations with a scalar field source, 
one of which is the the no-hair integral, defined in (\ref{condition1}), of the solution.
It is shown in \S\ref{sec_BH_NH_thms} that these criteria are satisfied for the solutions obtained in this work.
\par 
For a given spacetime
the presence of an event horizon can be inferred from 
analyzing 
 ingoing null trajectories, as explained in \S\ref{sec_sss}. 
A set of coordinates convenient for both tracking 
null trajectories and describing the whole neighborhood of an event horizon are 
Painlev\'e-Gullstrand (PG)  coordinates, originally
suggested in \cite{Painleve:21,Gullstrand:1922tfa}.
Below can be found a brief 
description of how they are derived and the manner 
in which PG coordinates describe null trajectories.
More complex spacetime tensors including the stress-energy and  Einstein tensors  in PG coordinates are given in \S\ref{sec_GPC}.
\par 
PG coordinates were originally suggested as an alternative to Schwarzschild coordinates for describing 
radial null trajectories in Schwarzschild spacetime.
The unique solution of the Einstein equation
that is spherically-symmetric, stationary, 
non-spinning with no net charge  is 
the Schwarzschild metric with the form 
\begin{equation}
ds^2=-fdt^2+\frac{1}{f}dr ^2+r^2d\Omega^2, \label{Schwarzschild}
\end{equation}
where $d\Omega^2=d\theta^2+\sin^2\theta d\varphi^2$ is the line-element of a unit two-sphere and $f=1-2M/r$ is the Schwarzschild 
term with $M$ being the mass of the background. 
The four-velocity $U^\mu=dx^\mu/d\tau \equiv \dot x^\mu$ 
($\tau$ being proper time along the worldline) 
satisfies the normalization condition 
$-1 = g_{\mu \nu} U^\mu U^\nu = -f\dot t^2 + f^{-1}\dot r^2 =U^2\left(-f+f^{-1}(dr/dt)^2\right)$
where $U\equiv dt/d\tau$.
The quantity  
$\varepsilon= -g_{\mu\nu}\xi^\mu U^\nu=fU$ 
is a constant of motion, since $\xi^\mu=(\partial /\partial t)^\mu$ is a timelike Killing field. 
Accordingly, the four velocity of a radially outgoing or ingoing spherically-symmetric worldline  is
\begin{equation}
U^\mu
= \left( \frac\varepsilon{f} \,,\, -\sqrt{\varepsilon^2-f},\, 0,\,0 \right).
\end{equation}
\par 
In PG coordinates, the time coordinate, denoted $t_\mathrm{p}$ to distinguish it from the $t$ in 
Schwarzschild coordinates, is the proper time along the worldline of the geodesic.
As such the four velocity now has the more natural form
\begin{equation}
U_{\rm p}^\mu=\left(\dot{\mathstrut t_{\mathrm{p}}},\,\dot{\mathstrut r},\,\dot{\mathstrut \theta},\,\dot{\mathstrut \phi}\right) 
= \left( 1 \,,\, -\sqrt{\varepsilon^2-f},\, 0,\,0 \right)\equiv (1,-v ,0,0)\end{equation}
where a `dot' refers to a derivative with respect to 
$t_{\rm p}$, and 
\begin{equation}
v =\sqrt{\varepsilon^2-f } 
\end{equation}
is the radial component of the velocity on the free-falling trajectory.
The PG time coordinate $t_{\rm p}$ is related to the Schwarzschild time coordinate as
\begin{equation}
d{t_\mathrm{p}} = \varepsilon d{t_\mathrm{s}} + \frac{\sqrt{\varepsilon^2-f}}{f}dr. 
\end{equation}
After writing the Schwarzschild metric (\ref{Schwarzschild})
in PG coordinates 
  the Painlev\'e-Gullstrand metric is obtained with the  form 
\begin{equation}
ds^2 = -{dt_\mathrm{p}}^2 +\frac{1}{\varepsilon^2}\left(dr+v {dt_\mathrm{p}}\right)^2 +r^2{d\Omega}^2. \label{gPG}
\end{equation}
Note that as pointed out in \cite{Kanai:2010ae}, the  form of the metric in (\ref{gPG})
is somewhat analogous to the conserved Newtonian energy  
\begin{equation}
E = \frac{1}{2}\left(\frac{dr}{dt_{\mathrm{p}}}\right)^2+\Phi(r)   \label{CL}
\end{equation}
where $\Phi(r)=-M/r$ is a Newtonian type potential and
$E=(\varepsilon^2-1)/2$ is constant.
If the particle  falls from rest at infinity,  $\varepsilon=1$, $E=0$, such that 
(\ref{gPG}) reduces to the standard  Painlev\'e Gullstrand metric
\begin{equation}
ds^2 = -{dt_\mathrm{p}}^2 +\left(dr+\sqrt{\frac{2M}{r}}d{t_\mathrm{p}}\right)^2 +r^2{d\Omega}^2. \label{PG}
\end{equation}
Both forms of the metric in (\ref{gPG}) and (\ref{PG}) are regular at the horizon $r=2M$,
ergo the spacetime geometry 
inside and outside the horizon of a black hole can be related without 
the emergence of any singularities.
As explained in \cite{Kanai:2010ae}, the Newtonian type energy motivates
the following ansatz for the metric in the generalized Painlev\'e-Gullstrand form: 
\begin{equation}
ds^2 = -{dt_\mathrm{p}}^2 +\frac{1}{1+2E(t_\mathrm{p},r)}\Bigl(dr+v(t_\mathrm{p},r){dt_\mathrm{p}}\Bigr)^2
+r^2{d\Omega}^2, \label{GPG}
\end{equation}
where 
\begin{equation}
v(t_\mathrm{p},r)=\sqrt{2E(t_\mathrm{p},r)+\frac{2m(t_\mathrm{p},r)}{r}}. 
\label{intro_PG_metric}
\end{equation}
Here  $E$ and $m$ are
not constant values, rather they are functions of $t_{\mathrm{p}}$ and $r$.
The metric in (\ref{intro_PG_metric}) is the one used in this paper, but
without an explicit dependence on  $t_{\mathrm{p}}$, namely only stationary solutions are considered.
\par 
The metric signature is taken to be $(-1,1,1,1)$,
and we use natural units, namely $c=\hbar=1$ is assumed.
In the opening section we defer setting the Newtonian constant $G=1$
for the purpose of offering clarity in our calculations,
but later it will be set to unity.
\par 
This paper is structured in the following way.
In section   \ref{sec_model_under_constructn} we introduce our model for $q$-theory,
which is effectively a scalar field theory with a double-well potential interaction, minimally coupled to Einstein gravity.
In section \ref{sec_sss}  the Einstein equations are given for the case of a static  spherically symmetric spacetime,
in two different sets of coordinates: generalized Painlev\'e-Gullstrand,  and those that shall be referred to as generalized
Schwarzschild coordinates. 
In section \ref{sec_BH_NH_thms}
the restrictions on solutions to the field equations are explain, which arise due to the no-hair theorems
that hold for 
scalar field theories minimally coupled to gravity.
Section \ref{sec_V} contains
 a detailed discussion of one specific static and spherically symmetric $q$-theory solution to the field equations.
Section  \ref{sec_VI}  builds on section  \ref{sec_V}, containing a local scan of the space of solutions around the solution considered in section  \ref{sec_V}. 
In section  \ref{sec_VII}   instabilities of the  obtained solutions are addressed,
both due to classical perturbations as well as due to Hawking radiation. 
We end our work in \ref{sec_VIII} with a conclusion of our findings.
\section{The model under consideration}\label{sec_model_under_constructn}\noindent
In this paper we consider a gravitating dynamical vacuum of the type introduced in Refs. \cite{Klinkhamer:2007pe,Klinkhamer:2008ns}. 
One way to describe such a system is through a  $3$-form field $A_{\beta\gamma\delta}$, antisymmetric with respect to permutation of indices. 
From a stand point, this system can be considered  to be matter described by a field $A_{\beta\gamma\delta}$, interacting with the gravitational field.   
The secondary scalar field $q$, which is the effective degree of freedom at low energies, is composed  of a three form field $A_{\beta\gamma\delta}$ as 
\begin{align}
&q^2=-\frac{1}{24}F_{\alpha\beta\gamma\delta}F^{\alpha\beta\gamma\delta},\,\, &&F_{\alpha\beta\gamma\delta}=\nabla_{[\alpha}A_{\beta\gamma\delta]},\\
&F_{\alpha\beta\gamma\delta}=\pm q\sqrt{-g}\varepsilon_{\alpha\beta\gamma\delta},\,\, &&F^{\alpha\beta\gamma\delta}=\pm q\frac{1}{\sqrt{-g}}\varepsilon^{\alpha\beta\gamma\delta}\ ,
\end{align}
where $\varepsilon_{\alpha\beta\gamma\delta}$ and $\varepsilon^{\alpha\beta\gamma\delta}$ are completely antisymmetric, namely
$\varepsilon_{0123}=1$ and $\varepsilon^{0123}=-1$. The square brackets denote antisymmetrization of indices. The freedom of choice of sign in relating $q$ with $F_{\alpha\beta\gamma\delta}$ is indicated by $\pm$ but will not be of further concern. From these relations it follows that
\begin{align}
\frac{\delta q}{\delta g^{\alpha\beta}}=\frac{1}{2}qg_{\alpha\beta}\ .
\label{qvariation}
\end{align}
\par 
The action of the model has the form
\begin{align}
S=\int d^4x\sqrt{-g}\left(\frac{R}{16\pi G}-\epsilon (q)-\frac{1}{2}g^{\alpha\beta}\nabla_{\alpha}q\nabla_{\beta}q\right)
\label{action1}
\end{align}
where $G$ is Newton's constant.
$\epsilon$ is a polynomial function in $q$ that has the form 
\begin{align}
\epsilon (q)=\frac{\lambda}{4}\left(q^4-\frac{1}{aG}q^2\right).
\end{align}
$\lambda$ and $a$ are real numbers assumed to be $O(1)$-parameters with $\lambda ,a>0$. The scale associated with the potential function is due to $G$ and, therefore, it is the Planck scale.\par 
Variation of the action with respect to the metric results in the Einstein equations:
\begin{align}
R_{\alpha\beta}-\frac{1}{2}g_{\alpha\beta}R=-8\pi G\left[g_{\alpha\beta}\left(\rho (q)+\frac{1}{2}\nabla_{\alpha}q\nabla^{\alpha}q\right)-\nabla_{\alpha}q\nabla_{\beta}q+2\Box q\frac{\delta q}{\delta g^{\alpha\beta}}\right]
\label{graviequation}
\end{align}
where 
\begin{align}
\rho (q)=\epsilon (q)-\frac{d\epsilon}{dq}\left(\frac{1}{2}g^{\mu\nu}\frac{\delta q}{\delta g^{\mu\nu}}\right)=\epsilon (q)-\frac{d\epsilon}{dq}q
\end{align}
 follows directly from (\ref{qvariation}). The function $\rho$ enters the Einstein equations in the same way as the cosmological constant. The shift from $\epsilon$ to $\rho$,
 as well as the final term on the right hand side of the Einstein equations, follow from the relation  $q=q(g_{\mu\nu})$. A self-sustained quantum vacuum fulfills
\begin{align}
0=P=-\rho
\end{align}
in thermodynamic equilibrium,
where $P$ refers to pressure  and $\rho$ is the energy density.
This yields, in our example, the equilibrium values $q_ {eq}=0,\pm \frac{1}{\sqrt{3aG}}$,
which satisfy $\rho (q_{eq})=0$. \mz{By further analogy with thermodynamics, an equilibrium chemical potential $\mu_{eq}$ may be defined as}
\mz{\begin{align}
\mu_{eq} =\frac{d\epsilon}{dq}\bigg\rvert_{q=q_{eq}}.
\label{chemicalpotential}
\end{align}}
Given that the original potential function is even with respect to $q$,
and assuming that the vacuum has a non-trivial equilibrium configuration, it shall be assumed that $q_{eq}=\frac{1}{\sqrt{3aG}}$ from now on. \mz{Therefore, $\mu_{eq} = - \frac{\lambda}{2(3 a G)^{3/2}} $. }
	
	Variation of the action with respect to the three - form field $A_{\alpha\beta\gamma}$ yields the generalized Maxwell equation
	\begin{align}
		&\nabla_{\alpha}(\sqrt{-g}\left(-\frac{d\epsilon (q)}{dq}\frac{\delta q}{\delta F_{\alpha\beta\gamma\delta}}+\Box q\frac{\delta q}{\delta F_{\alpha\beta\gamma\delta}})\right)=0\\
		&\Leftrightarrow \epsilon^{\alpha\beta\gamma\delta}\nabla_{\alpha}\left(-\frac{d\epsilon (q)}{dq}+\Box q\right)=0\\
		&\Leftrightarrow \frac{d\epsilon (q)}{dq}-\Box q=\mu\label{rhomu}
	\end{align}
	Here $\mu$ is the integration constant.
\mz{	Thus we obtain that the field equations for the $3$-form field $A_{\beta\gamma\delta}$ allow for an a priori free integration constant $\frac{d\epsilon (q)}{dq}-\Box q=\mu$ (which takes exactly the value of Eq. (\ref{chemicalpotential}) in the case of constant $q$-field). 
The energy density function inside the Einstein equations is then replaced by} 
\begin{align}
{\rho} (q)=\epsilon (q)-\mu q
\end{align}
\mz{with a constant $\mu$ that does not depend on coordinates. In the particular case, when the value of $\mu$ is equal to $\mu_{eq}$, the function ${\rho}(q)$ is reduced to } 
\begin{align}
	{\rho} (q)=\epsilon (q)-\mu_{eq} q=\frac{\lambda}{4}\left(q^4-\frac{1}{aG}q^2+\frac{2}{(3aG)^{\frac{3}{2}}}q\right)\, \label{rhomod}
\end{align}
It has a local minimum at $q=q_{\rm min}=-(\frac{1}{\sqrt{3}}+1)\frac{1}{\sqrt{4aG}}$, a local maximum at $q=q_{\rm max}=(-\frac{1}{\sqrt{3}}+1)\frac{1}{\sqrt{4aG}}$ and another local minimum at the equilibrium value $q=q_{eq}=\frac{1}{\sqrt{3aG}}$ with $\rho (q=q_{eq})=0$. Consequently, the equilibrium value is 
also a double root of $\rho$. The other roots are located at $q=0$ and $q_0=-\frac{2}{\sqrt{3aG}}$. \par 
Vacuum stability requires the vacuum compressibility $\chi_{\rm vac}$ to be positive :
\begin{align}
\chi_{\rm vac}^{-1}=\left(q^2\frac{d^2\epsilon }{dq^2}\right)\bigg|_{q=q_{eq}}=\frac{\lambda}{6a^2G^2}\geq 0\ ,
\end{align}
fulfilled for $\lambda ,a>0$.\par 
\begin{figure}
\begin{center}
\includegraphics[scale=0.4]{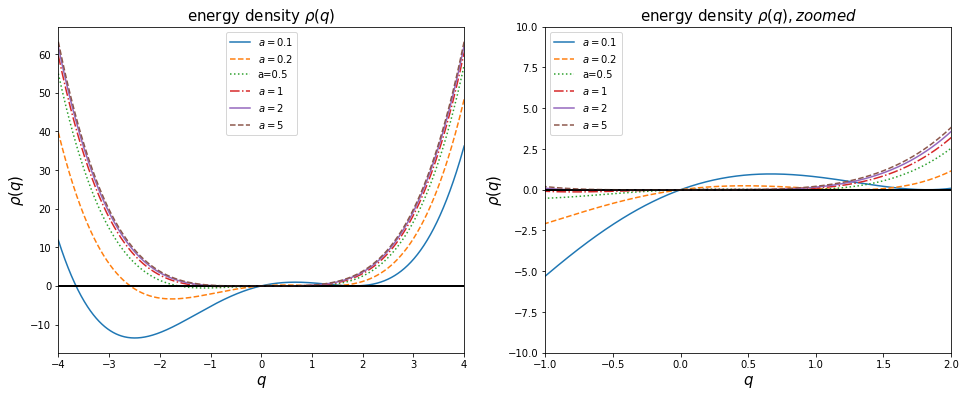}
\end{center}
\caption{The energy density of the q-field for $\lambda =1$ and different values of $a$ is depicted. While $\lambda$ sets the absolute scale, $a$ determines the depth and separation width of the potential wells.}
\label{edf}
\end{figure}
The energy density function for $\lambda =1$ and different values of $a$ is plotted in Fig. (\ref{edf}) with the black line marking the level of vanishing energy density. The potential 
has the same basic characteristic of two wells: the well containing the equilibrium value of the $q$-field as a minimum  with zero energy density, and the well that 
is deeper and hence allows for negative energy densities. The region of negative energy densities starts at $q=q_0$ and ends at $q=0$.\par
The action in (\ref{action1}) lacks higher derivative terms of the $q$-field, which usually appear in the effective field theory description without fine tuning. 
These  terms are (in the absence of an intermediate high-energy physics scale) suppressed by the Planck energy scale. As long as the $q$-field varies  slowly ($q^{\prime}\ll   1$ in units with $G=1$), these contributions are negligible. This is the approach adopted in the following part of the discussion.\par 
 Inserting \mz{Eq. (\ref{rhomu})} into Einstein equations (\ref{graviequation}) yields 
\begin{align}
R_{\alpha\beta}-\frac{1}{2}g_{\alpha\beta}R=-8\pi G\left[g_{\alpha\beta}(\epsilon (q)-\mu q+\frac{1}{2}\nabla_{\alpha}q\nabla^{\alpha}q)-\nabla_{\alpha}q\nabla_{\beta}q\right]
\label{modeinstein}
\end{align}
which comprises both the gravitational field and the matter (generalized Maxwell) equations. 
The fact that the $q$-field is not fundamental, but rather only an effective degree of freedom leads to the effective replacement of $\epsilon$ with $\rho$. 
In equilibrium, the scalar-field value corresponds to a minimum of the energy density $\rho$, which is  located at the value zero.\par
This shows that the problem reduces to that of solving the (modified) Einstein equations given by (\ref{modeinstein}). 
It is equivalent to a scalar field theory minimally coupled to gravity with a scalar field potential $\rho$.\par 
\section{Static and spherically symmetric solutions of the Einstein equations}\label{sec_sss}\noindent 
The discussion is now about solving  the static, spherically symmetric Einstein equations in order to find asymptotically flat solutions that describe a BH with a non-trivial $q$-field behavior. The $q$-field, as well as all the other functions, depend on a single coordinate  only, which we choose to be the standard radial coordinate. 
The q-field is expected to relax to the equilibrium value at large values of the radial coordinate, and to deviate from the equilibrium value 
approaching smaller and smaller values for the radial coordinate.\par 
In the following, two different ansätze are used for the metric in order to solve the Einstein equations:
\begin{align}
G^{\alpha}_{\,\,\,\,\beta}=8\pi G(T_q)^{\alpha}_{\,\,\,\,\beta},\,\, G^{\alpha}_{\,\,\,\,\beta}=R^{\alpha}_{\,\,\,\,\beta}-\frac{1}{2}g^{\alpha}_{\,\,\,\,\beta}R
\end{align}
where the Einstein tensor is $G^{\alpha}_{\,\,\,\,\beta}$ and the q-field energy-momentum tensor is $(T_q)^{\alpha}_{\,\,\,\, \beta}$.
Primes above symbols  label derivatives with respect to the radial variable, throughout this work.\par
\subsection{Generalized Painlev\'e coordinates}\label{sec_GPC}\noindent
The first ansatz is a generalized Painlev\'e-Gullstrand metric with the  form
\begin{equation}
ds^2=-dt^2+\frac{1}{1+2E(r)}(dr+v(r)dt)^2+r^2d\Omega^2
\end{equation}
and 
\begin{align}
v(r)=\sqrt{2E(r)+\frac{2Gm(r)}{r}}
\end{align}
while 
\begin{align}
d\Omega^2=d\theta^2+\sin^2(\theta )d\phi^2\ .
\end{align}
As explained in the introductory remarks, the terms $E$ and $v$
in the metric are related to the kinematical quantities of a particle in motion in the background. 
As elucidated in \cite{Kanai:2010ae}, $v(r)$ may be interpreted as the velocity of a freely falling test particle as it falls in towards  a (spherically symmetric) gravitating object from infinity, while $E(r)$,  at least asymptotically, can be related to the normalized total energy of a test particle ($E(\infty )=\frac{(e^2-1)}{2}$,
where $e$ represents the total energy per unit rest mass of a test particle at infinity). This motivates labeling $v$ a velocity function, and $E$ an energy function.\par
For the first metric ansatz, the non-vanishing Einstein tensor components are \cite{Kanai:2010ae}
\begin{align}
{G^t}_t=&-\frac{2Gm^{\prime}}{r^2}\ ,\label{Gtt}\\
{G^t}_{ r}=&\frac{-2E^{\prime}}{r(1+2E)}\sqrt{2E+\frac{2Gm}{r}}\ ,\label{Gtr}\\
{G^r}_r =&\frac{-2rE^{\prime}+4E^{\prime}Gm-4EGm^{\prime}-2Gm^{\prime}}{(1+2E)r^2}\ ,\\
{G^{\theta}}_\theta = {G^{\phi}}_\phi=&\frac1{(1+2E)^2 r^2} \left(3r^2\left(1-\frac{2Gm}{r}\right)(E^{\prime})^2+3r(1+2E)E^{\prime}Gm^{\prime}-(r+m)(1+2E)E^{\prime}
\right.\nonumber\\
&\left. -r^2\left(1-\frac{2Gm}{r}\right)(1+2E)E^{\prime\prime}-r(1+2E)^2Gm^{\prime\prime}\right)
\end{align}
The energy momentum tensor $(T_q)^{\alpha}_{\,\,\,\,\beta}$ for the $q$-field takes the form
\begin{align}
(T_q)^{\alpha}_{\,\,\,\,\beta}=-{g^{\alpha}}_{\beta}\left(\rho (q)+\frac{1}{2}g^{\mu\nu}\nabla_{\mu}q\nabla_{\nu}q\right)
-g^{\alpha\gamma}\nabla_{\gamma}q\nabla_{\beta}q
\end{align}
\mz{Here and henceforth we assume that the function $\rho$ is given by Eq. (\ref{rhomod}), i.e. with $\mu = \mu_{eq}$, which is mandatory for an asymptotically flat solution with $lim_{r\to\infty}q(r)=q_{eq}$. The $q$-field should satisfy the equation
\begin{align}
\mu	=\frac{d\epsilon (q)}{dq}-\Box q=lim_{r\to\infty}(\frac{d\epsilon (q)}{dq}-\Box q)=\frac{d\epsilon}{dq}\bigg\rvert_{q=q{eq}}=\mu_{eq}.
\label{qfieldeom}
\end{align}
That is 
\begin{equation}
	\frac{\lambda}{4} (4 q^3 - \frac{2}{aG}q)-\Box q=-\frac{\lambda}{2(3 a G)^{3/2}}.
\end{equation}
}
The non-vanishing components of stress - energy tensor are:
\begin{align}
&(T_q)^t_{\,\,\,\, t}=(T_q)^{\theta}_{\,\,\,\, \theta}=(T_q)^{\phi}_{\,\,\,\,\phi}=-\left(\rho (q)+\frac{1}{2}\left(1-\frac{2Gm}{r}\right)(q^{\prime})^2\right)\ ,\\
&(T_q)^t_{\,\,\,\, r}=\sqrt{2E+\frac{2Gm}{r}}(q^{\prime})^2\ ,\\
&(T_q)^r_{\,\,\,\, r}=-\rho (q)+\frac{1}{2}\left(1-\frac{2Gm}{r}\right)(q^{\prime})^2\ .
\end{align}
The Einstein equations can thus be brought into the following form
\begin{align}
&Gm^{\prime}=4\pi Gr^2\left(\rho (q)+\frac{1}{2}\left(1-\frac{2Gm}{r}\right)(q^{\prime})^2 \right)\label{einstein_eq_1}\\
&\frac{2E^{\prime}}{1+2E}=-8\pi Gr(q^{\prime})^2\label{einstein_eq_2}\\
&q^{\prime\prime}=\left(1-\frac{2Gm}{r}\right)^{-1}\left(-\frac{2q^{\prime}}{r}+\frac{2Gmq^{\prime}}{r^2}+8\pi Gr\rho q^{\prime}+\frac{d\rho}{dq}\right)\ .
\label{einsteingpg}
\end{align}
The first two equations are obtained from the radial and time components of the Einstein equations. Using these to simplify the equation due to the angular components leads to the third equation.\par
The second equation can be solved for $E$. With the definition $F=\ln(1+2E)$ and $E(\infty )=\lim\limits_{r\rightarrow\infty}E(r)$ the result is
\begin{align}
F(r)=\ln(1+2E(\infty ))+8\pi G\int_r^{\infty}s(q^{\prime}(s))^2\, ds\ .
\end{align}
Eqs.(\ref{einstein_eq_1}) and (\ref{einsteingpg}) can be solved numerically.\par
\subsection{Generalized Schwarzschild coordinates}\noindent
An analogous procedure for the second ansatz yields the metric in the form
\begin{align}
ds^2=-f(r)dt^2+\frac{1}{h(r)}dr^2+r^2d\Omega^2,
\label{metric2}
\end{align}
which shall be referred to as  generalized Schwarzschild coordinates.
From this form of the metric 
 the following non-vanishing components of the Einstein tensor are derived:
\begin{align}
&G^t_{t}=\frac{rh^{\prime}+h-1}{r^2},\qquad\qquad {G^r}_{r}=\frac{rhf^{\prime}+hf-f}{r^2f},\\
&{G^{\theta}}_{\theta}={G^{\phi}}_{\phi}=-\frac{1}{4rf^2}\left(rh(f^{\prime})^2-2rfhf^{\prime\prime}-2fhf^{\prime}-(rff^{\prime}+2f^2)h^{\prime}\right)
\end{align}
The non-vanishing components of the corresponding energy momentum tensor are
\begin{align}
(T_q)^t_{\,\,\,\, t}={(T_q)^{\theta}}_ \theta ={(T_q)^{\phi}} \phi =-\left(\rho (q)+\frac{1}{2}h(q^{\prime})^2\right)\ ,\qquad\qquad 
{(T_q)^r}_{r}=-\left(\rho (q)-\frac{1}{2}h(q^{\prime})^2\right).
\end{align}
The Einstein equations can finally be brought into the following form
\begin{align}
\label{einsteindiagonal1}
&1-h-rh^{\prime}=8\pi Gr^2(\rho +\frac{1}{2}h(q^{\prime})^2)\\
&\frac{f^{\prime}}{f}-\frac{h^{\prime}}{h}=8\pi G r(q^{\prime})^2\\
&q^{\prime\prime}=\frac{1}{h}\frac{d\rho}{dq}-\frac{h^{\prime}}{h}q^{\prime}-\frac{2}{r}q^{\prime}-4\pi Gr(q^{\prime})^3.
\label{einsteindiagonal3}
\end{align}
The first two equations are obtained from the radial and time components of the Einstein equations, and subsequently used to simplify the relation for the angular components. This results in the third equation.
The second equation can  be solved and reads, with $k=\ln(f)$ and the assumption $k(r=\infty )=0$, as  
\begin{align}
k(r)=-\int_r^{\infty}\left(\frac{h^{\prime}(s)}{h(s)}+8\pi Gs(q^{\prime}(s))^2)\right)\, ds\ .
\end{align}
The connection between the two parameterizations is provided by the relation
\begin{align}
h(r)=1-\frac{2Gm(r)}{r}
\end{align}
according to  which the first and third equations of the two ansätze are identical. For  convenience, and to be consistent with 
other literature on this topic,
 we introduce the notation 
\begin{align}
f(r)=h(r)e^{2\delta (r)}
\end{align}
and
\begin{align}
\tilde{\delta}(r_1,r_2)=\int_{r_1}^{r_2} 4\pi Gs \,q^{\prime}(s)^2\,ds\ .
\end{align}
\subsection{Black hole spacetime characteristics}\noindent
In \S\ref{sec_V} we will start from the assumptions that there exists an event horizon at a certain radial coordinate value $r=r_h$,  and  
that spacetime is asymptotically flat. 
It then  follows that
\begin{align}
\delta (r) = -\tilde{\delta}(r_h,\infty )+\tilde{\delta}(r_h,r),\,\,\,\, F(r)=\ln(1+2E(\infty ))+2\tilde{\delta}(r,\infty ).
\end{align}
In order to check for the existence of an event horizon, radial null geodesics in generalized Painlev\'e-Gullstrand coordinates or generalized Schwarzschild coordinates need to be discussed.
The requirement $ds^2|_{\theta =\theta_0,\phi=\phi_0}=0$ leads to 
\begin{align}
-dt^2+\frac{(dr+vdt)^2}{1+2E}=0 \Leftrightarrow \frac{dr}{dt}=\pm \sqrt{1+2E}-v
\end{align}
in  generalized Painlev\'e-Gullstrand coordinates and 
\begin{align}
-fdt^2+\frac{1}{h}dr^2=0 \Leftrightarrow \frac{dr}{dt}=\pm \sqrt{fh}
\label{gen_PG_coords}
\end{align}
in  generalized Schwarzschild coordinates. The plus sign is associated with outward motion, while the minus sign corresponds to inward motion. If $\frac{dr}{dt}<0$ (for both signs in (\ref{gen_PG_coords})), and for all $r<r_0$, then $r_0$ marks  the location of an event horizon. More precisely, it marks the location 
of an apparent horizon. In a static (or more generally in a stationary) spacetime,  the apparent horizon and the event horizon coincide. 
Generalized Schwarzschild coordinates are only valid on one side of the event horizon, since they become singular at an event horizon, in which case  either $f=0$ or $h=0$. In the vicinity of an event horizon, $h(r)\ll 1$, and the outward velocity of a massless particle may be approximated by
\begin{align}
\left(\frac{dr}{dt}\right)_{out}=&\sqrt{1+2E(r)}-v(r)\nonumber\\=&\sqrt{1+2E(r)}-\sqrt{2E(r)+\frac{2Gm}{r}}
\nonumber\\
=&\sqrt{1+2E(r)}-\sqrt{2E(r)+1-h(r)}\nonumber\\=&\frac{h(r)}{2\sqrt{1+2E(r)}}+O(h(r)^2).
\end{align}
As a result,  $r=r_h$ marks an event horizon if  $h(r_h)=0$. Regarding the Einstein equations in generalized Painlev\'e-Gullstrand ansatz given in  (\ref{einsteingpg}), it is clear that while the coordinates are regular on the horizon, the presence of a scalar field in Einstein gravity leads to singular behavior in the region approaching the 
horizon. 
This is clear also from the components of the Einstein equation in 
(\ref{einsteingpg})  of the first ansatz,
as well as from the components in Eq.~(\ref{einsteindiagonal3}) 
of the second ansatz.
A striking property of solutions  in  $q$-theory is the different distributions of energy 
inside and outside the event horizon.
Even more interesting  
is the question of singularity of solutions at the origin. \par
To answer these questions it is convenient to define
\begin{align}
-{(T_q)^t}_{t}=V(r)+T(r)\ ,\qquad V(r)=\rho (q(r))\ ,\qquad  T(r)=\frac{1}{2}h(r)(q^{\prime}(r)^2
\label{energydensity}
\end{align}
such that the energy density into a potential part $V$ and a kinetic part $T$.\par
The latter quantity may be deduced from the Kretschmann invariant 
\begin{align}
K(r)=R_{\mu\nu\rho\sigma}R^{\mu\nu\rho\sigma}
\end{align}
as $r\rightarrow 0$, where $R_{\mu\nu\rho\sigma}$ 
is the
Riemann 
curvature tensor. For generalized Schwarzschild coordinates it takes the explicit form
\begin{align}
\nonumber K^{{\rm q-theory}}(r)=
\frac{1}{r^4}\bigg[&4r^4h^2(r) \delta^{\prime}(r)^4+8r^4h^2(r)\delta^{\prime}(r)^2\delta^{\prime\prime}(r)+4r^4h^2(r) \delta^{\prime\prime}(r)^2\\
\nonumber &+8r^2h^2(r) \delta^{\prime}(r)^2+r^4h^{\prime\prime}(r)^2+\left(9r^4 \delta^{\prime}(r)^2+4r^2\right)h^{\prime}(r)^2\\
\nonumber &+4\left(3r^4h(r) \delta^{\prime}(r)^3+3r^4h(r)\delta^{\prime}(r)\delta^{\prime\prime}(r)+2r^2h(r)\delta^{\prime}(r)\right)h^{\prime}(r)\\
&+2\left(2r^4h(r) \delta^{\prime}(r)^2+2r^4h(r)\delta^{\prime\prime}(r)+3r^4\delta^{\prime}(r)h^{\prime}(r)\right)h^{\prime\prime}(r)+4(h(r)-1)^2\bigg].
\end{align}
For Schwarzschild spacetime with $h(r)=f(r)=1-\frac{2GM}{r}$ and Schwarzschild mass parameter $M$ the Kretschmann scalar reads
\begin{align}
K^{\rm Schwarz}(r)=\frac{48G^2M^2}{r^6}.
\end{align}
The existence of spacetimes minimally coupled to scalar fields
that contain BHs
depends 
on a number of conditions that follow from BH no-hair theorems. 
These criteria are discussed in the next chapter. 
Keeping explicit terms in $G$ in expressions is no longer necessary, hence 
the rest of the discussion will be in terms of geometric units, in which $G=1$.
\section{Black hole no-hair theorems}\label{sec_BH_NH_thms}\noindent
In this section the BH no-hair theorems are explained. 
Conditions arise from them that determine the existence of 
solutions of scalar hair black holes (SHBH's) in curved spacetime.
The class of solutions discussed in this work are 
themselves SHBH solutions. 
\par
The BH no-hair theorems, as discussed in  \cite{Heusler:1996ft,Sudarsky:1995zg}, assert the following:
\begin{enumerate}
\item  In the absence of event horizons there exist no non-trivial, regular scalar soliton solutions  that satisfy the 
dominant energy condition but violate the strong energy condition, at every point in an asymptotically flat spacetime.
\item
If the dominant energy condition holds but the strong energy condition does not,
in the presence of an event horizon in a static  spherically symmetric and asymptotically flat spacetime, no non-trivial 
regular scalar-field solution exists outside the event horizon, 
\item Any SHBH solution must necessarily have $V(r_h)<0$ where $r_h$ denotes the radial location of the event horizon, 
and $V$ is the potential energy density of the scalar field. The vicinity of the event horizon is enveloped by a region of negative scalar-field energy density.
\end{enumerate}
The strong energy condition is the requirement  that $T_{\alpha\beta}k^{\alpha}k^{\beta}\geq 0$,
where $T_{\alpha\beta}$ is the covariant energy momentum tensor and $k^{\alpha}$ is an arbitrary null vector field,
and   that $-T^{\alpha}_{\,\,\,\,\beta}p^{\beta}$ is a future pointing causal vector field whenever $p^{\alpha}$ is.
An equivalent way of stating the strong energy condition is that for every timelike vector field $u^{\alpha}$,
then  $(T_{\alpha\beta}-\frac{1}{2}T_{\mu\nu}g^{\mu\nu}g_{\alpha\beta})u^{\alpha}u^{\beta}\geq 0$.\par
If 
spherical symmetry is assumed with a positive scalar field potential,  the dominant energy condition does not need to be imposed 
in order to infer the absence of non-trivial scalar field solutions. From the first two criteria, it is immediate  that within the domain of outer communications (the region from the event horizon to asymptotically flat infinity), the scalar field energy density must have negative values. The third criterion 
then specifies where this region of negative energy density must be located. \par
This leaves as the only possibility for a non-trivial SHBH solution in the considered setup a $q$-field which asymptotically relaxes to its equilibrium value but sweeps over field values corresponding to negative potential energy density as the horizon is approached, for sure in the horizon proximity.\par
Both SHBH's, and scalar solitons, have to fulfill an integral equation as a necessary condition for existence,  as derived from a scaling argument \cite{Heusler:1996ft}. These conditions are written below in our notation convention.
A necessary condition for the existence of a scalar soliton (scalaron) in curved spacetime, in a non BH geometry is 
\begin{align}
\int_0^{\infty}4\pi r^2\exp\left(2\delta (r)\right)\left(E^{\rm flat}_{\rm kin}(r)+3V(r)\right)\,dr=0
\label{condition1}
\end{align}
It comprises the flat space kinetic energy density $E^{\rm flat}_{\rm kin}(r)=\frac{1}{2}(q^{\prime}(r))^2$,
as well as the potential energy density $V(r)=\rho (q(r))$. Analogously, a necessary condition for the existence of a SHBH solution is
\begin{align}
\int_{r_h}^{\infty}4\pi r^2\exp(2\delta (r))\left(\frac{2r_h}{r}\left(1-\frac{m(r)}{r}\right)-1\right)E^{\rm flat}_{\rm kin}(r)+\left(\frac{2r_h}{r}-3\right)V_{pot}(r))dr=0.
\label{condition2}
\end{align}
where $r_h$ is the radial coordinate of the event horizon. 
The fulfillment of this latter condition is taken as a tool to fine-tune the shooting parameter $q(r_h)$ (introduced and discussed in section V),
in finding SHBH solutions. In the limit $r_h\rightarrow 0$ (\ref{condition1}) is recovered. For future convenience we introduce the function
\begin{align}
nhf(r)=\int_{r_h}^{r}s^2\exp(2\delta (s))\left(\frac{2r_h}{s}(1-\frac{m(s)}{s})-1\right)E^{\rm flat}_{\rm kin}(s)+\left(\frac{2r_h}{s}-3\right)V(s))\,ds
\label{nhf}
\end{align}
which is then supposed to fulfill $\lim\limits_{r\rightarrow\infty}nhf(r)=0$.\par
\section{A representative SHBH solution for $q$-theory}\label{sec_V}\noindent
The existence of SHBH solutions has been known for some time and was first considered numerically in \cite{Nucamendi:1995ex} for a scalar field minimally coupled to gravity in a double well scalar field interaction potential. Subsequently, these solutions have been discussed in the framework of the 
of isolated horizons  \cite{Corichi:2005pa}. The latter formalism has been treated in detail in \cite{Ashtekar:2000nx} and references cited therein. The difference between this work and \cite{Nucamendi:1995ex} is in the more restrictive potential of $q$-theory. 
There are only two free parameters, the absolute scale provided by $\lambda$ and the depth or separation of the wells as parameterized by $a$. 
A shift in the $q$-field, accompanied by a corresponding change in the scalar energy density function, does not lead to a further quantitative change in the solution as induced by the shift in the $q$-field itself. It can be seen as fixed by the condition $\rho (q=0)=0$. The scalar potential in \cite{Nucamendi:1995ex} has three free parameters and allows for adjusting the positions of the wells, independently. Consequently, the solutions found and presented within that work cannot be used for $q$-theory, since they lie outside the parameter space spanned by $\lambda$ and $a$. In the following, we replace the parameter $a$ by the location of the minimum $q_{eq}=\frac{1}{\sqrt{3a}}$ of the shallow potential well.\par
We use Python for plotting and numerically solving the Einstein equations for the minimally coupled $q$-field. The solver is non-adaptive and makes use of a refined (fourth order) Runge-Kutta method. Refined means that the grid size close to the horizon is smaller than further away. This is due to the observation that the equations become singular at the horizon. We will nevertheless employ a prescription of how to start ``close'' to the horizon. \par
This comes about since only one of the three boundary conditions of the differential equations ($q(r=r_{\rm bound})$, $q^{\prime}(r=r_{\rm bound})$ and $m(r=r_{\rm bound})$ at radial coordinate boundary position $r=r_{\rm bound}$) is free for an asymptotically flat black hole spacetime regular at the horizon which we are searching for. This freedom resides in the radial coordinate location of the event horizon for a fixed scalar field potential function, as is the case for Schwarzschild spacetime. The dependencies of the boundary values on the horizon radius are known precisely only at the horizon except for one parameter, the ``shooting'' parameter. How the singular behavior at the horizon is circumvented is explained further below, together with an account of numerical errors. The results have been confirmed by an adaptive routine of the NDsolve-method in Mathematica, which is the most accurate. We will still stick with Python in the following discussion, since the difference between  the solving routines is negligibly small for tiny grid sizes (of the order $\sim 10^5-10^6$), as used in Python. \par
Both a discussion of how to numerically avoid the singularity in the third Einstein equation at the horizon, as well as a discussion of numerical errors are lacking in \cite{Nucamendi:1995ex}, but will be provided in the following. We shall start with a qualitative discussion of a representative solution in this section before focusing on the space of solutions in the next section.
\subsection{Outside the event horizon}\noindent
\begin{figure}
\begin{center}
\includegraphics[width=18cm, height=18cm]{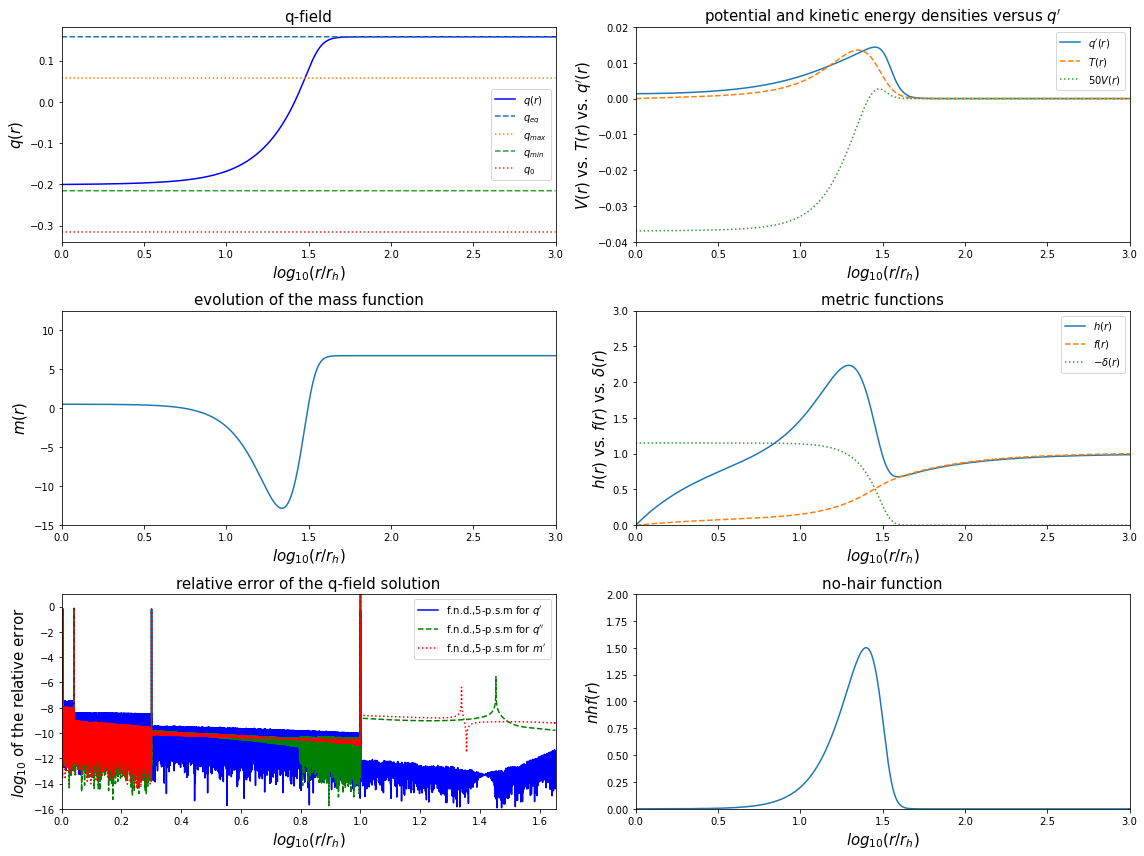}
\end{center}
\caption{A representative SHBH solution for $q$-theory outside the event horizon is shown for a horizon radius of $r_h=1$, $q$-field potential parameters $\lambda =1$ and $q_{eq}=0.158$, a grid size of $10^6$ and shift parameter $\epsilon =10^{-6}$.}
\label{ppssr7}
\end{figure}
The method of finding solutions for $q$-theory, makes use of previously discovered solutions, insofar as they are used as a starting point in an interpolation of solutions between the respective parameter spaces. For a horizon radius of $r_h=1$, $q$-field potential parameters $\lambda =1$ and $q_{eq}=0.158$, a grid size of $10^6$ and shift parameter $\epsilon =10^{-6}$ (to be introduced further below) the solution is shown in Fig. (\ref{ppssr7}) for the region outside the event horizon (also termed the domain of outer communications). It is in qualitative agreement with that of \cite{Nucamendi:1995ex}. \par
The $q$-field starts with a negative value such that $q_0<q(r=r_h)<0$, which is equivalent to $V(r)=\rho (q(r))<0$, as is required by the no-hair theorems. More specifically we have $q_{\rm min}<q(r=r_h)<0$. The $q$-field starts in the right half of the deep potential well shown in Fig. (\ref{edf}) with the characteristic points of the double well potential named as discussed in section I. It increases into the positive energy density domain, passes the local maximum and relaxes asymptotically to the equilibrium value $q_{eq}$. \par
The metric function $m(r)$, termed mass function, starts from the horizon with $m(r_h)=\frac{1}{2}r_h$, initially decreases as $\rho (q)<0$ until reaching a global minimum. It is located at a radial coordinate value shortly advancing that of $q=0$, since the kinetic energy density $T(r)=\frac{1}{2}h(r)(q^{\prime}(r))^2$ is positive everywhere outside the horizon. The mass function increases until finally approaching its asymptotic value, the ADM mass of the spacetime, from below. The potential and kinetic energy densities, as introduced in equation (\ref{energydensity}), are shown together with the derivative of the $q$-field on the upper right. The potential energy density is seen to be of minor size as compared to the kinetic part.\par
Further shown are the metric functions in generalized Schwarzschild coordinates in the central plot on the right hand side. As is required by asymptotic flatness we have the limits 
\begin{align}
\lim\limits_{r\rightarrow\infty}f(r)=1,\,\,\,\, \lim\limits_{r\rightarrow\infty}h(r)=1,\,\,\,\, \lim\limits_{r\rightarrow\infty}\delta (r)=0,
\end{align}
while $\lim\limits_{r\rightarrow\infty}E (r)$ remains a free parameter for the generalized Painlev\'e-Gullstrand coordinates. This free parameter is mostly irrelevant for the discussion of SHBH solutions and only of interest for test particle motion. Its variation will be discussed in Appendix A but is of no crucial importance elsewhere in the discussion. \par
The lowermost plot on the left shows the error sizes. We calculate the relative error of the solution by taking the first numerical derivative (f.n.d.) of $q$, $q^{\prime}$ and $m$ and comparing it with the solution as given by $q^{\prime}$, $q^{\prime\prime}$ and $m^{\prime}$ and obtained from the non-adaptive, refined (fourth order) Runge-Kutta algorithm. The first numerical derivative is calculated using a central $5$-point stencil method ($5$-p.s.m.). On a formal level this method approximates the derivative of a five times differentiable function accurately up to $O((\Delta r)^4)$-corrections where $\Delta r$ represents the grid spacing. The grid size is $10^6$. The first four peaks mark the radial coordinate values where the grid size changes abruptly and are artificial, since the $5$-p.s.m. formula we employed is based on equal grid size spacing. The final peaks for the relative errors are due to the appearance of the local extrema of $q^{\prime}$ as well as $m$. Apart from these points the relative error can be seen to be smaller than $10^{-7}$ throughout. \par
The lowermost plot on the right shows the function $nhf(r)$ introduced in the last section in equation (\ref{nhf}) which indeed fulfills $\lim\limits_{r\rightarrow\infty}nhf(r)=0$. This latter property was taken to fine-tune the shooting parameter $q(r_h)$, though, which will be discussed shortly.\par
It can be seen that the plot of the relative errors do not reach as far out as the other plots. An asymptotic analysis of the third Einstein equation with $\delta q(r)=q(r)-q_{eq}$ yields the linear approximation 
\begin{align}
(\delta q)^{\prime\prime}=-\frac{2(\delta q)^{\prime}}{r}+\rho^{\prime}=-\frac{2(\delta q)^{\prime}}{r}+\frac{\lambda}{2a}\delta q\ ,
\label{linearization}
\end{align}
which has the solution
\begin{align}
\delta q(r)=c_1\frac{1}{r}\exp\left(-\sqrt{\frac{\lambda}{2a}}r\right)+c_2\frac{1}{r}\exp\left(+\sqrt{\frac{\lambda}{2a}}r\right)\ .
\end{align}
The asymptotic behavior of the relevant functions can be deduced from the first Einstein equation and the condition $\lim\limits_{r\rightarrow\infty}m(r)<\infty$. A first order Taylor expansion of the potential energy density and its derivatives around $q=q_{eq}$ is well justified for $\delta q \ll   \frac{1}{\sqrt{a}}$. The approximation of the third Einstein equation is well justified for $\frac{2m(r)}{r},4\pi r^2\rho (q(r))\ll   1$. Taken together we then obtain (\ref{linearization}). The asymptotic solution has an exponentially growing and an exponentially depleting part.\par
In the search of a solution it is found that when the $q$-field and the mass function approach their asymptotic values they leave them again after some critical value. This can not be avoided and is an artifact of the numerical approximation. It induces a non-vanishing coefficient $c_2$ and therefore exponential growth due to numerical uncertainty. Therefore the exponentially depleting part is fitted onto the functions $q$, $q^{\prime}$ and $m$ after they tend to change very slowly by approaching their asymptotic values. This fitting of the free parameter $c_1$ takes place at those radial coordinate values where the relative error plots end. At this point $\frac{\delta q(r)}{q_{eq}},\frac{m(\infty )-m(r)}{m(\infty )},nhf(r)\ll   1$ hold.\par 
We now turn to the near horizon region. We search for solutions regular at the horizon by imposing $\lim\limits_{r\rightarrow r_h}q^{\prime\prime}(r)<\infty$. The third Einstein equation as well as the assumptions of the existence of an event horizon and of asymptotic flatness then reduce the freedom of the boundary conditions of $q$, $q^{\prime}$ and $m$ to one parameter. This parameter may be declared as the horizon radius. After fixing the scalar field potential parameters and thereby the theory, the space of solutions is one dimensional and parameterized by $r_h$ as is Schwarzschild spacetime. Up to one degree of freedom, the boundary values are known at the horizon. The event horizon condition $h(r)=0$ on the one hand and $\lim\limits_{r\rightarrow r_h}q^{\prime\prime}(r)<\infty$ on the other hand imply
\begin{align}
m(r_h)=\frac{r_h}{2},\,\,\,\, q^{\prime}(r_h)=\frac{d\rho}{dq}(q(r_h))\frac{r_h}{(1-8\pi r_h^2\rho (q(r_h)))}\ .
\label{boundaryconditionshorizon}
\end{align}
The remaining freedom then resides in the value of $q(r_h)$. It is adjusted so as to yield ("shoot" towards) an asymptotically flat solution and therefore termed shooting parameter. By choosing it more and more accurately the approach of $q$, $q^{\prime}$ and $m$ to their asymptotic values may be improved. This suggests that there exists exactly one shooting parameter which is appropriate for ensuring asymptotic flatness. We can only approach it to within a certain numerical accuracy. As soon as the solutions to the first and third Einstein equations are obtained, the horizon values of $E(r)$ and $\delta (r)$ may be deduced by integration of the second Einstein equation. They are given by 
\begin{align}
E(r_h)=((1+2E(\infty ))\exp(\int_{r_h}^{\infty}r(q^{\prime}(r))^2dr)-1)/2,\,\,\,\, \delta (r_h)=-\tilde{\delta}(r_h,\infty ).
\end{align}
The singular behaviour of the third Einstein equation at $r=r_h$ is avoided by the prescription $r\rightarrow r(1+i\epsilon )$. We call $\epsilon$ the shift parameter and choose it such that $0<\epsilon \ll   1$. The quantities plotted in Fig. (\ref{ppssr7}) are then understood as the real parts of the functions of the (total, complex valued) solution. The accuracy of the numerical calculations due to finiteness of grid size as well as shift parameter is analyzed in Appendix B.\par 
\subsection{Inside the event horizon}\noindent
\begin{figure}
\begin{center}
\includegraphics[width=18cm, height=18cm]{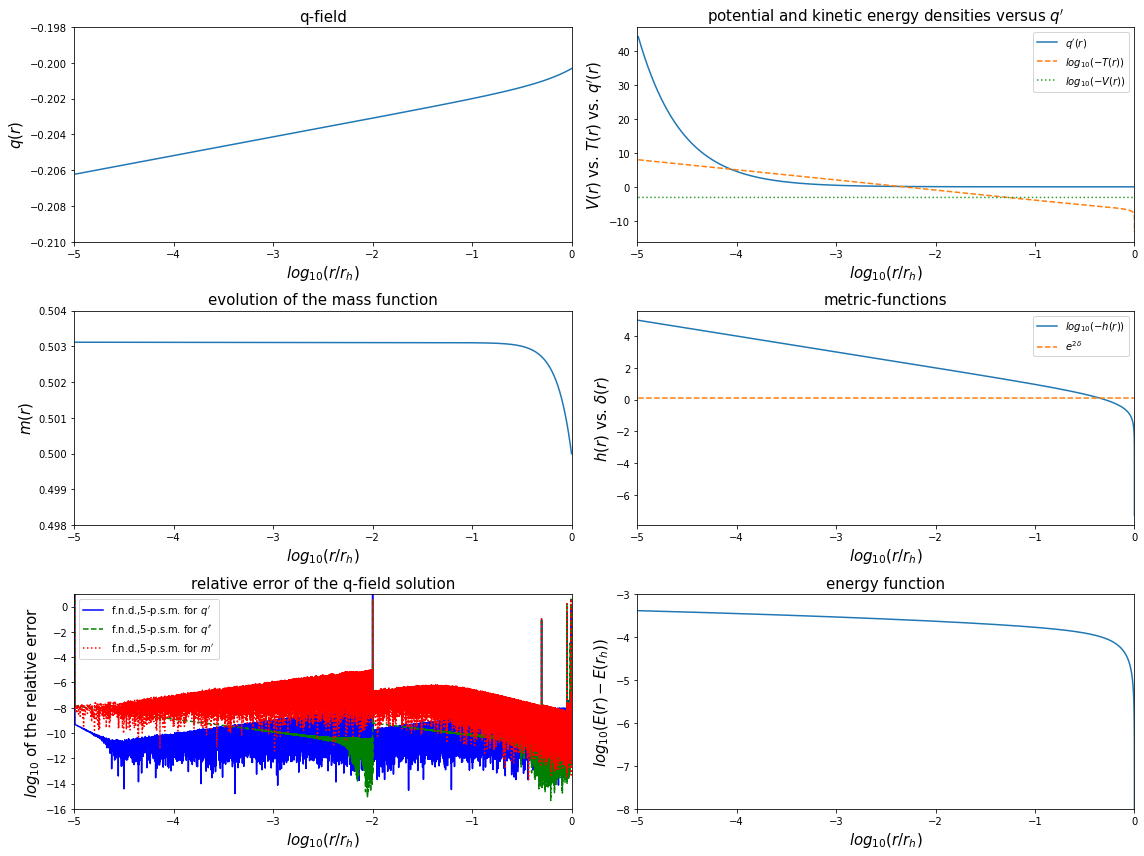}
\end{center}
\caption{A representative SHBH solution for $q$-theory inside the event horizon is shown for a horizon radius of $r_h=1$, $q$-field potential parameters $\lambda =1$ and $q_{eq}=0.158$, a grid size of $10^6$ and shift parameter $\epsilon =10^{-6}$.}
\label{ppssr9}
\end{figure}
In extension of the plots of Fig. \ref{ppssr7}, the solution for a horizon radius of $r_h=1$, $q$-field potential parameters $\lambda =1$ and $q_{eq}=0.158$, a grid size of $10^6$ and shift parameter $\epsilon =10^{-6}$ is shown in Fig. \ref{ppssr9} for the region inside the event horizon.
In distinction to the corresponding figure for the outside region the lowermost plot on the right hand side highlights the energy function. The relative error size remains below $10^{-5}$ for all of the functions $q^{\prime}$, $q^{\prime\prime}$ and $m^{\prime}$. The peaks mark again those radial coordinate values where the grid size changes abruptly. Close to the event horizon it has been chosen smaller as in the case of the region outside the event horizon. Of interest is the behaviour in the limit where the radial coordinate tends to zero. The solution is found to be singular in the $q$-field in this limit. The mass function, in contrary, tends to a constant value of about $\lim\limits_{r\rightarrow 0}m(r)=0.503$. An expansion of the third Einstein equation for $r\ll   1$ yields the approximate equation
\begin{align}
q^{\prime\prime}=-\frac{q^{\prime}}{r}
\end{align}
with solution
\begin{align}
q(r)=d_1\log_{10}(r)+d_2.
\end{align}
It is found that $d_1=0.0010\pm 0.0001$ and $d_2=-0.2010\pm 0.0001$ by a straight line fit. This implies the approximations
\begin{align}
F(r)&\approx ln(1+2E(r_h ))+8\pi\int_{r_0}^{r_h}r(q^{\prime}(r))^2dr+8\pi d_1^2(ln(r_0)-ln(r))\\
\nonumber &=C+8\pi d_1^2ln(\frac{1}{r}) \,\,\,\, \Leftrightarrow E(r)\approx \frac{\exp(C)}{2}r^{-8\pi d_1^2}-\frac{1}{2}\\ 
\delta (r)&\approx \delta (r_h)-\int_{r_0}^{r_h}4\pi r(q^{\prime}(r))^2dr-4\pi d_1^2(ln(r_0)-ln(r))\\
\nonumber &=D-4\pi d_1^2ln(\frac{1}{r}) \Leftrightarrow \exp(2\delta (r))\approx \exp(2D)r^{8\pi d_1^2},\\
h(r)&\approx 1-\frac{2m(0)}{r}
\end{align}
for $r<r_0\ll   1$ with for the further discussion irrelevant constants $C$ and $D$. Consequently the limits 
\begin{align}
\lim\limits_{r\rightarrow 0}q(r)=-\infty,\,\,\,\, \lim\limits_{r\rightarrow 0}q^{\prime}(r)=\infty 
\end{align}
for the $q$-field and its derivative and
\begin{align}
\lim\limits_{r\rightarrow 0}E(r)=\infty ,\,\,\,\, \lim\limits_{r\rightarrow 0}h(r)=-\infty ,\,\,\,\, \lim\limits_{r\rightarrow 0}\delta (r)=-\infty
\end{align}
for the metric functions follow. Since $d_1\ll   \frac{1}{\sqrt{8\pi}}$, $E(r)$ and $\exp(2\delta (r))$ vary very slowly as compared to $h(r)$. Therefore the behavior of the metric as $r\rightarrow 0$ will asymptote that of Schwarzschild spacetime. Especially, very close to the center the energy density will change sign again and becomes positive. The metric can not be continued to the radial coordinate origin.
\begin{figure}
\begin{center}
\includegraphics[scale=0.4]{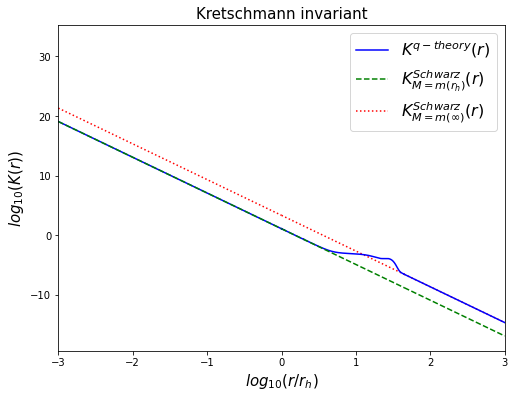}
\end{center}
\caption{The Kretschmann invariant for a representative SHBH solution for q-theory inside and outside the event horizon is shown for a horizon radius of $r_h=1$, $q$-field potential parameters $\lambda =1$ and $q_{eq}=0.158$, a grid size of $10^6$ and shift parameter $\epsilon =10^{-6}$. \mz{For comparison we represent the result for the Schwarzschild solutions corresponding to two different masses.}}
\label{ppssr14}
\end{figure}
There is a curvature singularity in the limit $r\rightarrow 0$ as is well known for Schwarzschild spacetime. {\mz{The Kretschmann invariants for $q$-theory and Schwarzschild spacetime for two different Schwarzschild mass parameters ($M=m(r_h)$ and $M=m(\infty )$) are shown in Fig. (\ref{ppssr14}). All of them show the same $\frac{1}{r^6}$-divergence as $r\rightarrow 0$. The transition of the Kretschmann invariant of $q$-theory between those of Schwarzschild spacetime arises in the region of radii where the $q$-field changes significantly. In this transition region the mass function changes from around $m(r_h)=\frac{r_h}{2}$ to $m(\infty )>m(r_h)$. The characteristics of the curve shape are consequences of the functional dependencies of the Kretschmann invariant on $m(r)$ (or equivalently $h(r)$) and its derivatives in this region. For vanishingly small radii ($r<<r_h$) the Kretschmann invariant in $q$-theory asymptotes to that of Schwarzschild spacetime with Schwarzschild mass parameter $M=m(r_h)=\frac{r_h}{2}$, while for very large radii ($r>>r_h$) the $q$-theory Kretschmann invariant asymptotes to that of Schwarzschild spacetime with Schwarzschild mass parameter $M=m(\infty )$.}}\par 
After discussing one solution in detail we proceed with a local scan of the space of solutions around that just presented. Both grid size and shift parameter will no longer be mentioned from now on and chosen to be very large in the former case and negligibly small in the latter as has been done within this section.
\section{The parameter space of SHBH solutions for $q$-theory}\label{sec_VI}\noindent
The qualitative features of the representative solution presented in the previous section are common to all SHBH solutions in q-theory. The space of solutions is parametrized by the horizon radius $r_h$ as well as the $q$-field potential parameters $\lambda$ and $q_{eq}$ (or as well $a$). In order to understand the differences between solutions we perform a local scan around the representative solution in the space of solutions and extract different quantities for each individual solution, in part in analogy with  \cite{Nucamendi:1995ex,Corichi:2005pa}.\par 
The ADM mass of a (static and spherically symmetric) solution is given by
\begin{align}
M_{\rm ADM}(r_h,\lambda ,q_{eq})=\lim\limits_{r\rightarrow\infty}m_{r_h,\lambda ,q_{eq}}(r).
\end{align}
We decompose it into two contributions 
\begin{align}
M_{\rm ADM}(r_h,\lambda ,q_{eq})=M_{\rm ADM}^{Schwarz}(r_h)+M_{hair}(r_h,\lambda ,q_{eq}).
\end{align}
The first contribution is the mass parameter of a Schwarzschild black hole of horizon radius $r_h$, $M_{\rm ADM}^{Schwarz}(r_h)=\frac{r_h}{2}$. It is independent of the scalar field potential parameters, as it describes a (static and spherically symmetric) black hole in vacuum. The dressing by the non-constant $q$-field outside the event horizon yields the non-vanishing additional contribution 
\begin{align}
M_{hair}(r_h,\lambda ,q_{eq})=&-\int_{r_h}^{\infty}4\pi r^2T^t_{\,\,\,\, t}dr\\
=&\int_{r_h}^{\infty}4\pi r^2(\rho (q(r)) +\frac{1}{2}h(r)(q^{\prime}(r))^2)dr.
\end{align}\par 
Further quantities of interest are the shooting parameter as well as the radial coordinate location relative to $r_h$ and absolute depth of the global minimum of the mass function as functions of the parameters of the space of solutions. The relative radial coordinate location of the global minimum of the mass function is important insofar as it marks the region where both the $q$-field and the mass function vary significantly.
\subsection{Dependence on the horizon radius}\noindent
\begin{figure}
\begin{center}
\includegraphics[width=18cm, height=18cm]{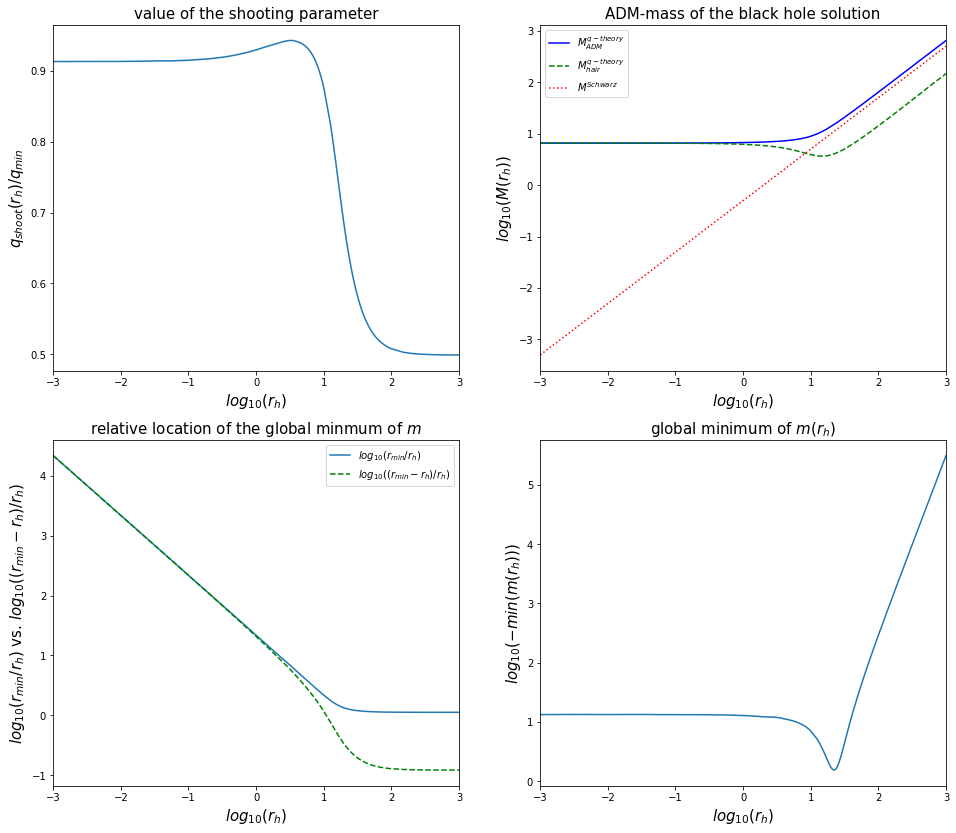}
\end{center}
\caption{The variation of several characteristic quantities of SHBH solutions for $q$-theory with respect to the horizon radius $r_h$ is shown for the $q$-field potential parameters $\lambda =1$ and $q_{eq}=0.158$. These quantities include shooting parameter, ADM mass, scalar-hair mass, relative location of the global minimum of the mass function as well as the absolute value of the global minimum of the mass function.}
\label{ppssr1}
\end{figure}
Fig. (\ref{ppssr1}) illustrates the characteristic quantities introduced above for a horizon radius range $10^{-3}<r_h<10^3$ and scalar field potential parameters coincident with those of the representative solution presented in section V. The lower limit has been chosen such that the asymptotic dependence on $r_h$ is visible. The upper bound is due to lack of precision of fitting the asymptotic part of the $q$-field and mass function. The asymptotic plateau becomes difficult to identify, since both $q$-field and mass function behave less and less smooth in the fitting region. The asymptotics for large horizon radii seem to be deducible from the plots as well, though. We will assume that the plots show the true asymptotics in both extreme situations $r_h\ll   1$ and $r_h\gg   1$.\par 
We may then draw the following conclusions:
\begin{itemize}
\item[1)] The shooting parameter $q_{\rm shoot}(r_h)=q(r_h)$ varies significantly. Its asymptotic values are $\lim\limits_{r_h\rightarrow 0}q_{\rm shoot}(r_h)=0.913q_{\rm min}$ and $\lim\limits_{r_h\rightarrow\infty}q_{\rm shoot}(r_h)=0.500q_{\rm min}$, respectively.
\item[2)] For $r_h\ll   1$ the ADM mass and scalar hair mass converge towards the value\par 
$\lim\limits_{r_h\rightarrow 0}\log_{10}(M_{\rm ADM}(r_h)),\, \log_{10}(M_{scalar \,\, hair}(r_h))=0.817$. For $r_h\gg   1$ both ADM mass and scalar hair mass increase linearly with the horizon radius and may be parameterized by 
\begin{align}
&\log_{10}(M_{\rm ADM}(r_h))=c_1\log_{10}(r_h)+c_2\\
&\log_{10}(M_{scalar \,\, hair}(r_h))=c_3\log_{10}(r_h)+c_4
\end{align}
with $c_1=1.0031\pm 0.0001$, $c_2=-0.1985\pm 0.0001$, $c_3=1.0136\pm 0.0001$ and $c_4=-0.8747\pm 0.0001$ obtained by a straight line fit. The $q$-theory ADM mass is a monotonically increasing function of the horizon radius and everywhere larger than the corresponding Schwarzschild spacetime ADM mass (which coincides with the mass parameter).
\item[3)] The region outside the event horizon where the $q$-field and mass function vary significantly approaches the event horizon relative to its size for $r_h\gg   1$. So it seems that in this regime the $q$-field behaves non-trivially only just outside the horizon and relaxes to its equilibrium value very quickly in the near horizon regime. This region of significant change does not approach the horizon infinitely close but relaxes to the value \par 
$\lim\limits_{r_h\rightarrow\infty}r_{\rm min}(r_h)/r_h=1+e^{-0.916}=1.400$. \par 
In the opposite limit $r_h\ll   1$ the region of significant change of both $q$-field and mass function gets pushed further and further away from the horizon region . {\mz{The functions in the lower left plot of Fig. (\ref{ppssr1}) have been extrapolated as far as $log_{10}(r_h)=-6$ an remain in accordance with the asymptotics implied by the shown range of the horizon radius. This implies that for the very limit $r_h\rightarrow 0$ no scalar soliton (scalaron) exists (contrary to the conclusions drawn in \cite{Nucamendi:1995ex}, which apply for a different locus in their extended parameter space outside of our range of parameters, though).}} Rather the $q$-field remains constant outside the event horizon with a value of $q_{\rm shoot}(0)=\lim\limits_{r_h\rightarrow 0}q_{\rm shoot}(r_h)=0.913q_{\rm min}$. The corresponding energy density is negative resulting in a Schwarzschild anti de Sitter spacetime with cosmological constant $\Lambda=8\pi G\rho (q_{\rm shoot}(0))$. 
\item[4)] The size of the global minimum of the mass function converges to a constant for $r_h\ll   1$ with value $\lim\limits_{r_h\rightarrow 0}\log_{10}(-{\rm min}(m)(r_h))=1.125$. For $r_h\gg   1$ it decreases linearly and faster than the negative ADM mass. It may be parametrized by 
\begin{align}
\log_{10}((-{\rm min}(m)(r_h))=d_1\log_{10}(r_h)+d_2
\end{align}
with $d_1=3.0268\pm 0.0001$ and $d_2=-3.5790\pm 0.0002$.
\end{itemize}
\begin{figure}
\begin{center}
\includegraphics[width=18cm, height=18cm]{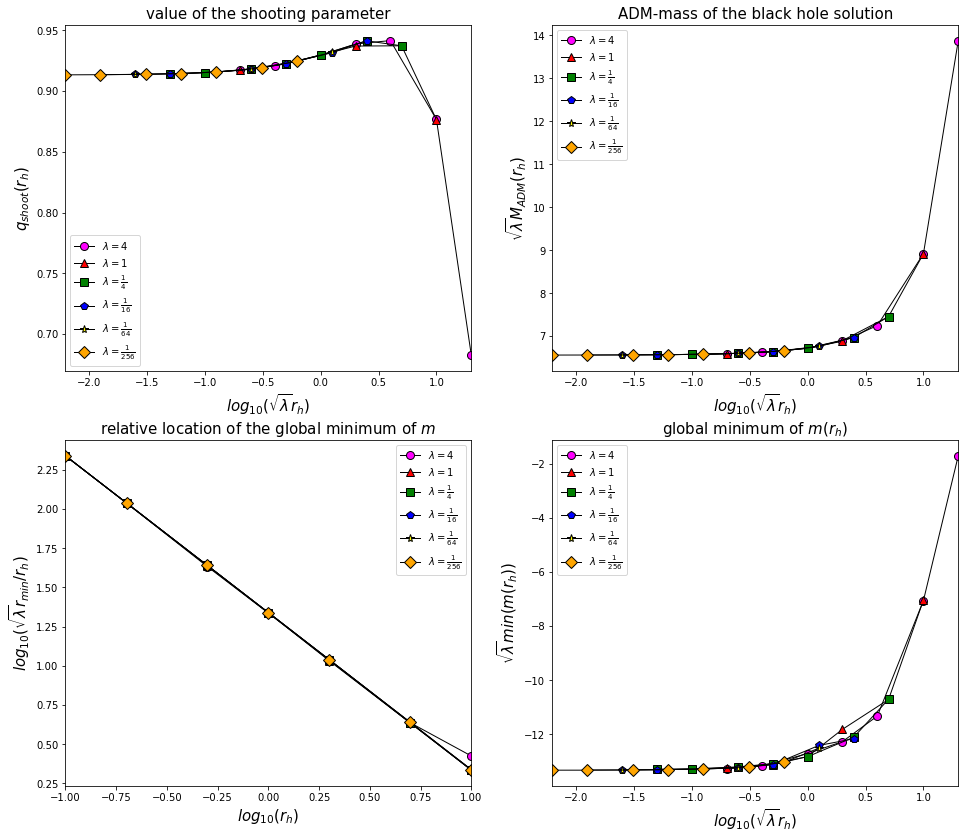}
\end{center}
\caption{The variation of several characteristic quantities of SHBH solutions for $q$-theory with respect to the horizon radius $r_h$ and several discrete values of the scalar field potential parameter $\lambda$ is shown with the remaining $q$-field potential parameter fixed at $q_{eq}=0.158$. These quantities include shooting parameter, ADM mass, scalar-hair mass, relative location of the global minimum of the mass function as well as absolute value of the global minimum of the mass function.}
\label{ppssr3}
\end{figure}
\subsection{Dependence on the parameters of the scalar field potential}\noindent
We now consider changes in the scalar field potential parameters.
The effect of a variation of the scalar field potential parameter $\lambda$ for a horizon radius range $\frac{1}{10}\leq r_h\leq 10$ and fixed scalar field potential parameter $q_{eq}=0.158$ for the characteristic functions presented previously in Fig. (\ref{ppssr1}) is visualized in Fig. (\ref{ppssr3}). The radial parameters and masses have been rescaled in a particular way following \cite{Nucamendi:1995ex}. It can be seen that the rescaled quantities depend only on two instead of three parameters (at least in the considered region of parameter space). To be more precise about the parameter dependencies define 
\begin{align}
\widehat{r_h}=\sqrt{\lambda}r_h, \,\,\,\,\widehat{r_{\rm min}}=\sqrt{\lambda}r_{\rm min},\,\,\,\hat{M}_{ADM}=\sqrt{\lambda}M_{\rm ADM},\,\,\,\,\widehat{{\rm min}(m)}=\sqrt{\lambda}{\rm min}(m).
\end{align}
It then follows that
\begin{align}
\widehat{r_{\rm min}}=1,\,\,\,\,\hat{M}_{ADM}=\hat{M}_{ADM}(\widehat{r_h},q_{eq}),\,\,\,\,\widehat{{\rm min}(m)}=\widehat{{\rm min}(m)}(\widehat{r_h},q_{eq}).
\end{align}
\begin{figure}
\begin{center}
\includegraphics[width=18cm, height=18cm]{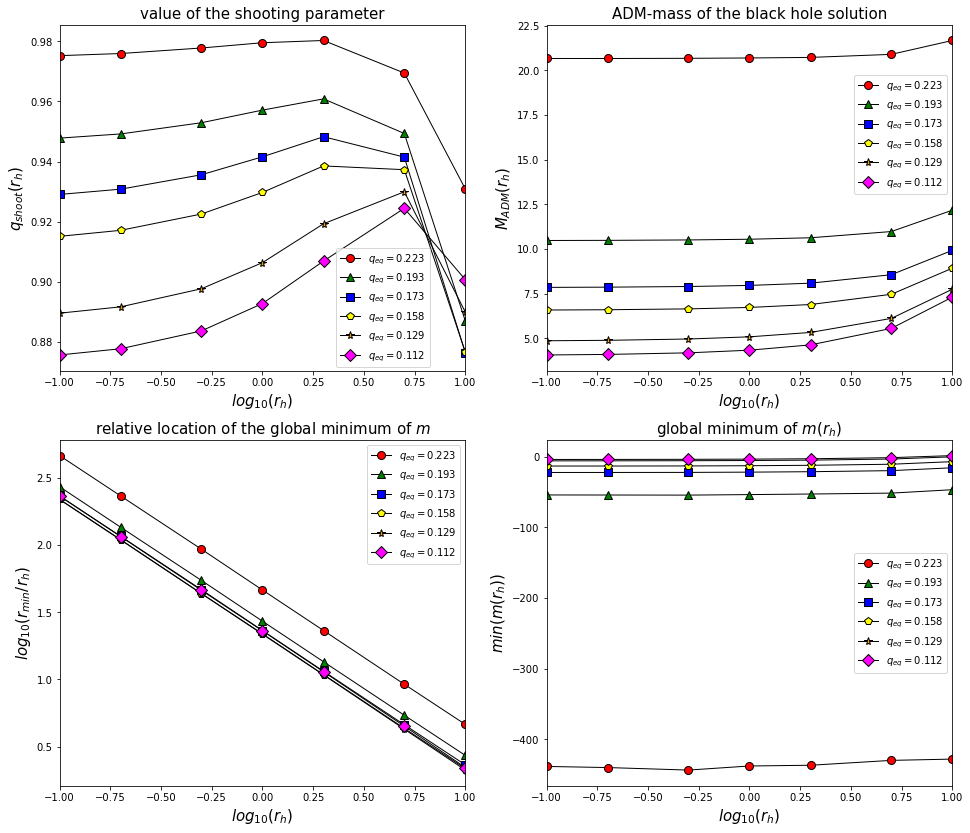}
\end{center}
\caption{The variation of several characteristic quantities of SHBH solutions for $q$-theory with respect to the horizon radius $r_h$ and several discrete values of the scalar field potential parameter $q_{eq}$ is shown with the remaining $q$-field potential parameter fixed at $\lambda =1$. These quantities include shooting parameter, ADM mass, scalar-hair mass, relative location of the global minimum of the mass function as well as absolute value of the global minimum of the mass function.}
\label{ppssr4}
\end{figure}
The effect of a variation of the scalar field potential parameter $q_{eq}$ for a horizon radius range $\frac{1}{10}\leq r_h\leq 10$ and fixed scalar field potential parameter $\lambda =1$ for the characteristic functions presented previously is visualized in Fig. (\ref{ppssr4}). As $q_{eq}$ increases, the potential wells of the scalar field potential recede from each other and become more pronounced. All quantities (with minor exceptions) seem to be monotonically growing (monotonically decreasing in the case of the negative valued quantity ${\rm min}(m)$) with $q_{eq}$. The dependence of the different characteristic quantities of SHBH solutions on $q_{eq}$ is not so easily deducible. Nevertheless it seems that, with exception of large horizon radius values, the dependence on $q_{eq}$ may be factorized from that on $\widehat{r_h}$.\par 
The local scan of the parameter space of solutions has revealed that the space of solutions is effectively only two dimensional with the dependence on the two parameters factorizing by a monotonically increasing function of $q_{eq}$ to good approximation within the represented parameter space area. \par 
The topic of the following section will be a stability analysis of SHBH solutions due to perturbations of the $q$-field solution.
\section{Stability of the SHBH solutions}\label{sec_VII}\noindent
An important question to ask is whether SHBHs are stable. Do perturbation modes of the $q$-field and the metric which grow exponentially in the SHBH spacetime of our $q$-theory model exist? The question of classical instability has been discussed for spherically symmetric, time dependent perturbations of the metric and scalar field (here the $q$-field) in 
\cite{Nucamendi:1995ex}. Following a similar notation to that introduced in \cite{Nucamendi:1995ex} (see Eqs. (17)-(19) therein) we define the s-wave perturbations by
\begin{align}
\tilde{q}(t,r)&=q(r)+\delta q(r,t)\ ,\label{pert_1}\\
\tilde{f}(r,t)&=f(r)(1-h_1(r,t))\ ,\label{pert_2}\\
\tilde{h}(r,t)&=h(r)(1-h_2(r,t))\  \label{pert_3}
\end{align} 
in generalized Schwarzschild coordinates. The functions $q(r)$, $f(r)$ and $h(r)$ represent the unperturbed solutions for $q$-theory
in the generalized Schwarzschild coordinates, while $\delta q(r,t)$, $h_1(r,t)$ and $h_2(r,t)$ denote small perturbations to the non-perturbed solution. 
Note that in Eq.~(\ref{pert_3}) 
$h_2$ is defined as a small correction to $h$ with   $g_{rr}=1/h$
in (\ref{metric2}), whereas in Eq.~(19) in 
\cite{Nucamendi:1995ex} it is defined as a small correction to $g_{rr}$ directly
with a different sign. The two relations are equivalent, as is easily checked by substituting
$h(1-h_2)$ for $h$ in $g_{rr}=1/h$ and then expanding.
\par 
After linearization of the Einstein equations (\ref{einsteindiagonal1}-\ref{einsteindiagonal3}) it can be shown that
\cite{Nucamendi:1995ex} 
\begin{align}
\partial_r h_1(r,t)=\partial_{r}h_2(r,t)-16\pi r\left(\,\partial_r\, q(r)\ \right)\,\left(\partial_r\,\delta q(r)\right)\ ,\qquad h_2(r,t)=8\pi r\,\left(\,\partial_r\, q(r)\ \right)\,\delta q(r).
\end{align}\noindent
As such the metric perturbations are expressible in terms of the $q$-field perturbation. 
The latter may be determined by solving a one-dimensional Schrödinger equation of the form \cite{Nucamendi:1995ex} 
\begin{align}
\left(\frac1{2m} (-i\partial_{r_\ast})^2 +V^\ast_{\rm eff}(r_\ast)\right)\psi (r_\ast)=\frac{1}{2m}(i\partial_t)^2 \psi (r_*),\qquad 
\frac{dr_\ast(r)}{dr}=\frac{\exp(-\delta (r))}{h(r)}
\end{align}
where $\psi (r_*(r))\equiv r\delta q(r)$ and $r_*$ is the ``tortoise'' coordinate. The scalar field mass $m$ is obtained as follows. Insert $\tilde{q}(t,r)$ into the action functional and expand in $\delta q(t,r)$ around $q(r)$. Far away from the event horizon when $q(r)$ is close to its equilibrium value $q_{eq}$, the kinetic cross terms as well as the linear potential term in $\delta q$ are negligible and the action with dynamical field (perturbation) $\delta q$ has a proper kinetic term with at least quadratic potential terms. The quadratic term yields $m$ in the same way as does the real Klein Gordon action with (self-)interactions. The so obtained value for $m$ reads
\begin{equation}m=\left.\sqrt{\dfrac{d^2\rho (q)}{dq^2}}\right|_{q=q_{\rm eq}}=\sqrt{\dfrac{3}{2}\lambda q_{\rm eq}^2}.\end{equation}
It will formally not be needed in the following but has been introduced in order to provide a properly normalized Schrödinger problem.
The effective potential reads 
\cite{Nucamendi:1995ex} 
(note that the expression for $V_{\rm eff}$ in this work is defined with a factor of $\frac1{2m}$ compared to Eq. (22) in \cite{Nucamendi:1995ex})
\begin{align}
\nonumber V_{eff}(r_*(r))=\frac{1}{2m}h(r)\exp(2\delta (r))[&\frac{h(r)}{r}(\delta^{\prime}(r)+\frac{h^{\prime}(r)}{h})\\
\nonumber &-8\pi r h(r)(q^{\prime}(r))^2(\delta^{\prime}(r)+\frac{h^{\prime}(r)}{h(r)}+\frac{1}{r})\\
&+16\pi rq^{\prime}(r)\frac{d\rho}{dq}(q(r))+\frac{d^2\rho }{dq^2}(q(r))].
\end{align}
The separation ansatz $\psi (r_*(r))=\xi \left(r_*(r)\right)\exp\left(\pm i\sqrt{2mE}t\right)$ 
leads to the following stationary Schrödinger equation for $\xi \left(r_*(r)\right)$
\begin{align}
\left(\frac{1}{2m}(-i\partial_{r_\ast})^2 +V^\ast_{\rm eff}(r_\ast)\right)\xi (r_\ast)=E\xi (r_\ast)
\label{schroedinger}
\end{align}
with energy eigenvalue $E$. A sufficient condition for static, spherically symmetric configurations to be unstable 
\cite{Wald:1992bd} is that the differential operator on the left-hand side of (\ref{schroedinger}),
is negative in the Hilbert space $L^2 (M)$, where $M$ is the spacetime manifold on which the metric is defined.
Accordingly, a sufficient condition for unstable solutions is the existence of a bound state $E<0$ in the Schrödinger problem.
In the same manner the existence of bound states that correspond to $E<0$ in the Schrödinger problem with potential $V_{\rm eff}$
is equivalent to the existence of unstable perturbations of
$q$-theory solutions.
The exponential factor of the perturbation becomes of order unity after a time of order 
\begin{align}
\tau = \frac{1}{\sqrt{-2mE_{\rm min}}}.\label{tau}
\end{align}
The subscript $\rm min$ on $E$ signifies the lowest bound state energy which is of most importance for giving the scale of $\tau$ (for a discrete and finite bound state spectrum, as is the case here). 
This time can be seen as the lifetime of the SHBH in the presence of these perturbations. 
\begin{figure}
\begin{center}
\includegraphics[width=12cm,height=10cm]{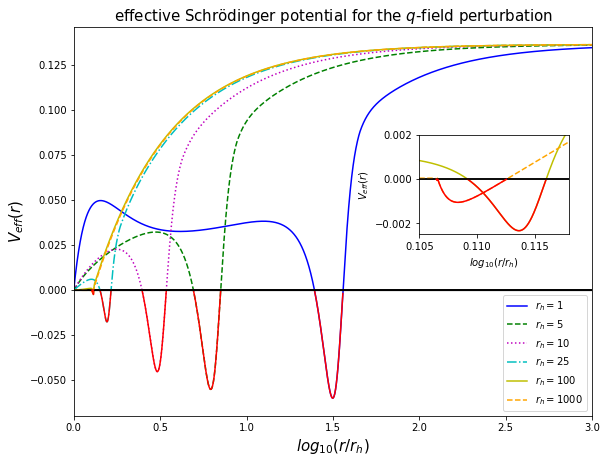}
\end{center}
\caption{The effective potential of the scalar perturbation mode $\xi (r_*)$ is shown for different horizon radii and scalar field potential parameters $\lambda =1$ and $q_{eq}=0.158$. It comprises a negative valued well where the wave function of bound states are predominantly located. The well is highlighted in red. A zoom makes this region visible for the horizon radii $r_h=100$ and $r_h=1000$.}
\label{ppssr12}
\end{figure}
\par 
In order to determine the value $E_{\rm min}$, 
the lowest eigenvalue of bound state energies of the effective Schrödinger equation (\ref{schroedinger}),
we solve the eigenvalue problem numerically in the variable $r$ for horizon radii in the range $1\leq r_h\leq 1000$ by approximating the equation on a grid of finite size. The differential operator is approximated by a central point stencil method accurate to fourth order of the grid spacing. The Runge Kutta solver shares this level of accuracy. We employ vanishing boundary conditions for the eigenfunctions in the limits $r\rightarrow 0$ as well as $r\rightarrow\infty$. To get an impression of the shape of the eigenfunction to the eigenvalue $E_{\rm min}$,
we replace the effective potential, illustrated in Fig. (\ref{ppssr12}) for different horizon radii, by an auxiliary potential. 
This auxiliary potential is a parabola fitted to the negative potential well of the effective potential in the region shown in red in the figure (where $V_{\rm eff}<0$). Outside it is set to zero. This yields a finite depth harmonic oscillator potential. The solutions of the wave equation may be determined exactly in this auxiliary potential. Finding the bound state energy eigenvalues is then identical to the quantum harmonic oscillator except for them being finite in amount, while the eigenfunctions are compromised due to the vanishing of the potential. Nevertheless, as is in concordance with the results in \cite{Nucamendi:1995ex}, the expectation is that the lowest energy bound state eigenfunction is close to a Gaussian in shape which is the exact eigenfunction in this case for the quantum harmonic oscillator.
\begin{figure}
\begin{center}
\includegraphics[width=12cm,height=10cm]{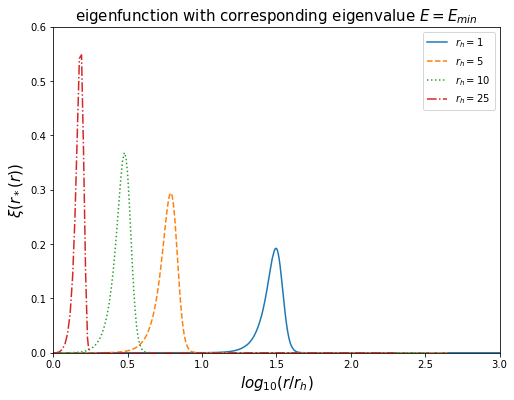}
\end{center}
\caption{Normalized eigenfunction solutions to the lowest bound state eigenvalue for the scalar perturbation mode $\xi (r_*)$ are shown for different horizon radii and scalar field potential parameters $\lambda =1$ and $q_{eq}=0.158$. They are almost perfect Gaussians in shape which is the case for the quantum harmonic oscillator.}
\label{ppssr16one}
\end{figure}
That this expectation is also fulfilled in our case is illustrated in Fig. (\ref{ppssr16one}). We plot the normalized eigenfunctions corresponding to the lowest bound state energy eigenvalues for several horizon radii and scalar field potential parameters $\lambda =1$ and $q_{eq}=0.158$. A comparison with Fig. (\ref{ppssr12}) shows that the peaks of the eigenfunctions are situated almost exactly at the minimum of the effective Schrödinger potential. It becomes apparent here and has been observed that for increasing horizon radii the Gaussians loose their shape and tend to disappear. In this regime the lowest eigenvalues approach zero very quickly from below. This inspires the conclusion that large SHBHs are indeed stable. {\mz{In \cite{Nucamendi:1995ex} instability is claimed, but only for two specific solutions within the space of solutions. This is therefore not in contradiction to our findings.}}
\begin{figure}
\begin{center}
\includegraphics[width=12cm,height=10cm]{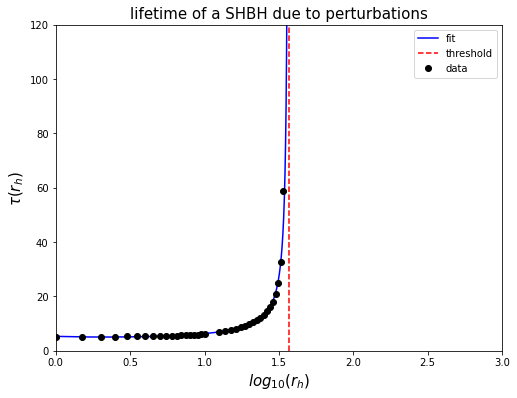}
\end{center}
\caption{The lifetime of SHBHs due to s-wave $q$-field and metric perturbations represented by the scalar perturbation mode $\xi (r_*)$ is shown as a function of the horizon radius with scalar field potential parameters $\lambda =1$ and $q_{eq}=0.158$. The lifetime is finite for small SHBHs below a certain radius or mass threshold value, whereas SHBHs are stable beyond this threshold. The threshold is highlighted in red.}
\label{ppssr16two}
\end{figure}
In Fig. (\ref{ppssr16two}) the lifetime $\tau =\tau (r_h)$ of SHBHs as a function of the horizon radius and scalar field potential parameters $\lambda =1$ and $q_{eq}=0.158$ is shown. We fit the obtained lifetimes $\tau =\tau (r_h)$ corresponding to the solutions for the lowest eigenvalues $e=e(r_h)$ to the function
\begin{align}
f(r_h)=a\cdot (\log_{10}(r_h))^ntan(c\cdot \log_{10}(r_h)-b)+d
\end{align}
parameterized by the scale parameters $a$ and $c$, the horizontal shift parameter $b$, the vertical shift parameter $d$ and the power parameter $n$ with optimal parameters $p_{opt}$ and covariance matrix $p_{cov}$ given by 
\begin{align}
p_{opt}=
\begin{pmatrix}
a\\
b\\
c\\
d\\
n
\end{pmatrix}=
\begin{pmatrix}
1.53\\
1.14\\
1.73\\
5.31\\
1.23
\end{pmatrix},\,\,\,\,
p_{cov}=\begin{pmatrix}
0.044 & 0.082 & 0.053 & 0.037 & -0.007\\
0.081 & 0.179 & 0.115 & 0.085 & 0.014\\
0.053 & 0.115 & 0.074 & 0.055 & 0.009\\
0.037 & 0.085 & 0.055 & 0.043 & 0.010\\
-0.007 & 0.014 & 0.009 & 0.010 & 0.028
\end{pmatrix}.
\end{align}
The lifetime then becomes infinite at the finite horizon radius $r^0_h=\frac{\pi +2b}{2c}$ and the SHBHs are therefore stable beyond the threshold $r^0_h$ for the chosen scalar field potential parameters.\par 
The stability analysis has revealed that SHBH are unstable due to classical s-wave perturbations of both the $q$-field and the metric below a certain size or equivalently mass threshold and stable beyond (at least for the chosen parameters). \par 
\section{Conclusion}\label{sec_VIII}\noindent
In the present paper we consider $q$-theory comprising a scalar field $q$ minimally coupled to gravity. 
The $q$-field describes a dynamical gravitating vacuum. \mz{As suggested by Volovik}  \cite{Volovik:2021qvc},
this theory may contain BH solutions that resemble that of a gravastar, i.e. a configuration with energy concentrated inside a thin spherical shell. \mz{Contrary to the conventional gravastar of \cite{Mazur:2001fv} (according to \cite{Volovik:2021qvc} this is the type I gravastar), the state proposed in \cite{Volovik:2021qvc} (i.e. the type II gravastar) contains an event horizon. Using  direct numerical calculations, we confirm the existence of similar BH configurations, with some reservations, though. Namely, inside the event horizon space - time resembles the interior of the Schwarzschild BH, and does not contain the de Sitter - like domain. Besides, the mentioned thin shell is located outside the event horizon, not inside. In these respects our configuration differs from those of \cite{Mazur:2001fv} and \cite{Volovik:2021qvc}. It may, therefore, be referred to as the type III gravastar.} Furthermore, there should exist a region in space, where the energy density is negative. This is required to satisfy the ``no-hair'' theorems. As a result, the thin shell situated just outside of the horizon contains both a piece of negative energy and a piece with positive energy density. The integration of the energy density inside the shell yields a total positive energy resulting in the ADM mass perceived by the distant observer.\par 
According to Eq. (\ref{Gtt}) the energy density is proportional to the derivative of the mass function $m(r)$. The latter function is represented within Fig. \ref{ppssr7}. One can see that for the given example solution, the spherical shell of finite thickness exists and is situated just outside the horizon. Inside this shell the essential variation of $m(r)$ is localized. Close to but outside the event horizon, the energy density is negative, then it passes through zero, and becomes positive in the second piece of the shell. The shell ends where $m(r)$ exponentially approaches its asymptotic value, the constant that represents the black hole mass seen by the infinitely distant observer.\par 
The results of section \ref{sec_VI} demonstrate that the spherical shell approaches the horizon relative to its size when the horizon radius is increased. In the limit of very large BHs, virtually the entire energy due to the $q$-field is localized in the thin shell situated  outside the horizon and close to it. It does not approach the event horizon ifinitely close, but stops, such that the energy density sign transition is  positioned around $\frac{7}{5}r_h$, where $r_h$ denotes the event horizon radius.\par
\mz{Notice that we do not consider in the present paper ordinary matter. This suggests the extension of the present work to the system of gravitating vacuum and conventional matter. This, however, remains out of the scope of the present paper, and at the present moment we cannot point out definitely the region of space, where conventional matter may be concentrated being added to the considered system.}
 
A stability analysis with respect to the s-wave metric and $q$-field perturbations shows that the BH solutions of the type considered in the present paper may be classically unstable . However, the corresponding configurations are stable for sufficiently large BHs. We therefore claim that stable heavy SHBHs do exist, {\mz{at least in a spherically symmetric setup}}.\par 
We do not discuss here  questions related to the stability of the considered configurations on the quantum level. This issue remains outside of the scope of the present paper.\par 
In conclusion, in this work we confirm the supposition of \cite{Volovik:2021qvc} about the existence of BH solutions in $q$-theory that look similar to  gravastars.  These states escape the conditions of the no-hair theorem, due to the region in space with negative energy density. At spatial infinity these solutions approach the Schwarzschild solution, but differ from it essentially close to the horizon. Inside the horizon, the vacuum density is negative and changes sign very close to the center. The singularity of curvature at $r=0$ is the same as that of the Schwarzschild solution. {\mz{The singularity at the center can perhaps be removed by an extended version of $q$-theory allowing for a dynamical, $q$-field dependent gravitational constant. This possibility arises naturally within the $q$-theory framework and has been discussed in \cite{Klinkhamer:2008ns}}}.\par 
The authors are grateful to G.E.Volovik for the proposition to consider the given problem, and for useful discussions during the initial stage of the work.        
\begin{appendix}
\section*{Appendix A. Test particle characteristics.}\par 
\begin{figure}
	\begin{center}
		\includegraphics[scale=0.3]{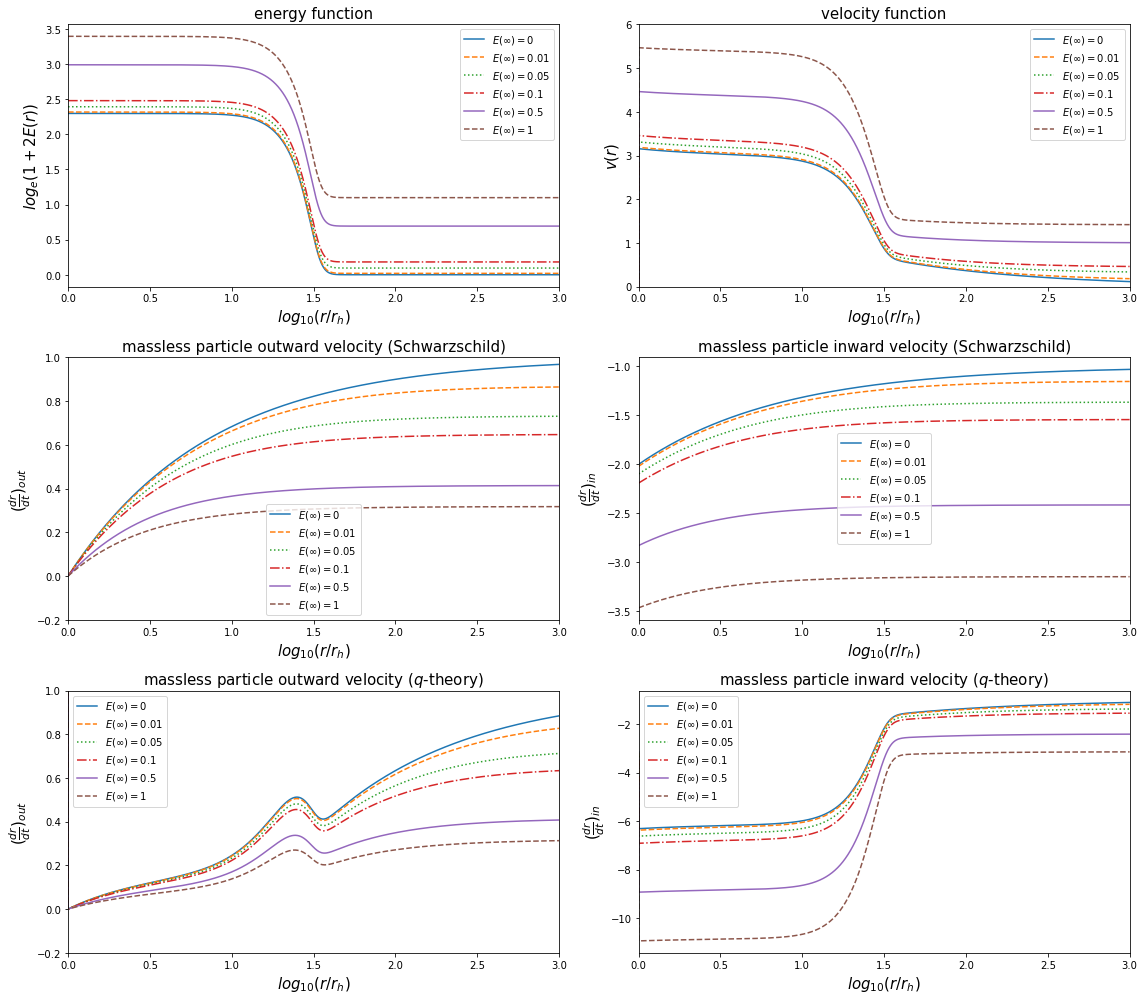}
	\end{center}
	\caption{The effect of different choices of $E(\infty )$ on the SHBH solution for q-theory outside the event horizon is shown for a horizon radius of $r_h=1$, $q$-field potential parameters $\lambda =1$ and $q_{eq}=0.158$, a grid size of $10^6$ and shift parameter $\epsilon =10^{-6}$. The inward and outward massless particle velocities are shown for both Schwarzschild spacetime and $q$-theory spacetime with Schwarzschild mass parameter $M=m^{q-theory}(r_h)$.}
	\label{ppssr8}
\end{figure}
\end{appendix}
We discuss here the properties of test particles contained in the free parameter $E(\infty )=\lim\limits_{r\rightarrow\infty}E(r)$. It is related to the total energy per unit rest mass of a test particle $e$ by $E(\infty )=(e^2-1)/2$ which moves towards the SHBH horizon starting at infinity with initial velocity $v(\infty )=\sqrt{2E(\infty )}$ where $v(\infty )=\lim\limits_{r\rightarrow\infty}v(r)$. Different values for $E(\infty )$ correspond to different initial kinetic energies of a test particle. The choice $E(\infty )=0$ corresponds to a particle at rest, while $E(\infty )>0$ corresponds to a particle initially moving towards the SHBH. $E(\infty )<0$ is not possible, since then $v(r)$ would become imaginary while approaching asymptotically flat infinity. The motivation for the introduction of generalized Painlev\'e-Gullstrand coordinates as well as their relation to test particle motion are presented in more detail in \cite{Kanai:2010ae}.\\
The numerical solution presented in Fig. \ref{ppssr7} for different initial values of the free parameter $E(\infty )$ is shown in Fig. \ref{ppssr8}. The function $E(r)$ is monotonically decreasing with $r$ as is $v(r)$ in $q$-theory, whereas $E(r)$ is constant for Schwarzschild spacetime while $v(r)$ is also decreasing. This is in concordance with the increase of the kinetic energy of a test particle as it moves towards the event horizon of the black hole. The inward and outward velocities of a massless particle $(\frac{dr}{dt})_{in/out}$ are monotonically increasing with $r$, as the gravitational pull of the black hole decreases by further recession from the horizon. This is true for both Schwarzschild and $q$-theory spacetime with one exception in $q$-theory. In the region of large change of the $q$-field the velocity of outward moving massless particles has a small dip before increasing again. In this region the energy density of the $q$-field becomes positive. As expected, the outward motion tends to zero as the event horizon is approached. We chose the Schwarzschild mass parameter $M=m^{q-theory}(r_h)$. Choosing $M=m^{q-theory}(\infty )$ instead would have implied that the graphs for $(\frac{dr}{dt})_{out}$ in the Schwarzschild case cross zero already outside the event horizon as set by $q$-theory.  Close to the event horizon both inward and outward massless particle velocities are smaller for $q$-theory spacetime as compared to Schwarzschild spacetime. This suggests that black hole absorptivity is greater for $q$-theory spacetime as compared to Schwarzschild spacetime.\\

\section*{Appendix B. Accuracy estimates.}
\begin{figure}
	\begin{center}
		\includegraphics[scale=0.3]{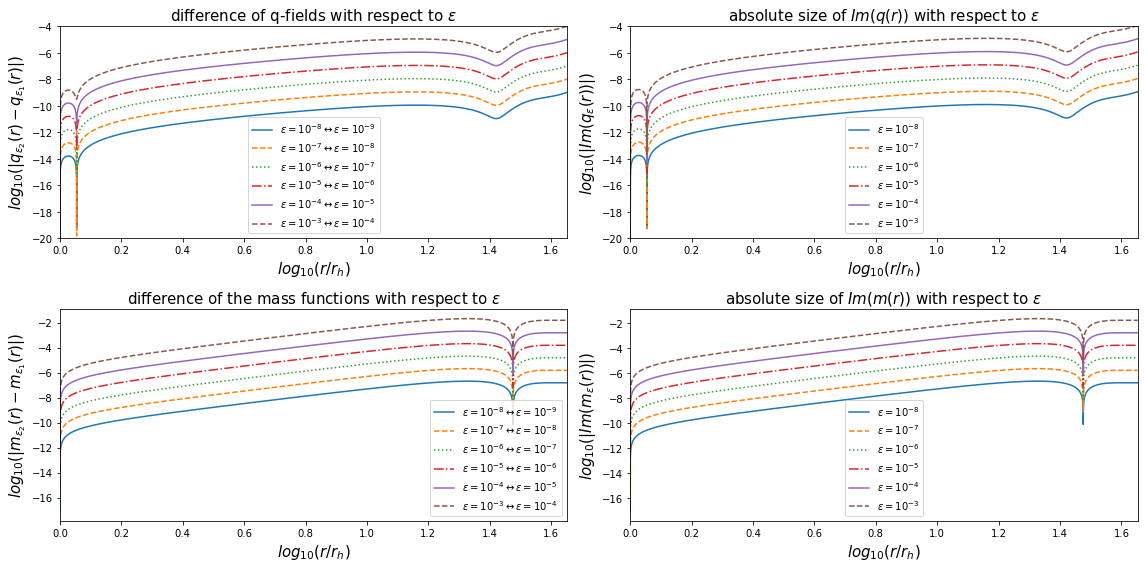}
	\end{center}
	\caption{The effect of different shift parameters on SHBH solutions for $q$-theory outside the event horizon is shown for a horizon radius of $r_h=1$, $q$-field potential parameters $\lambda =1$ and $q_{eq}=0.158$ and a grid size of $10^6$.}
	\label{ppssr5}
\end{figure}
\begin{figure}
	\begin{center}
		\includegraphics[scale=0.3]{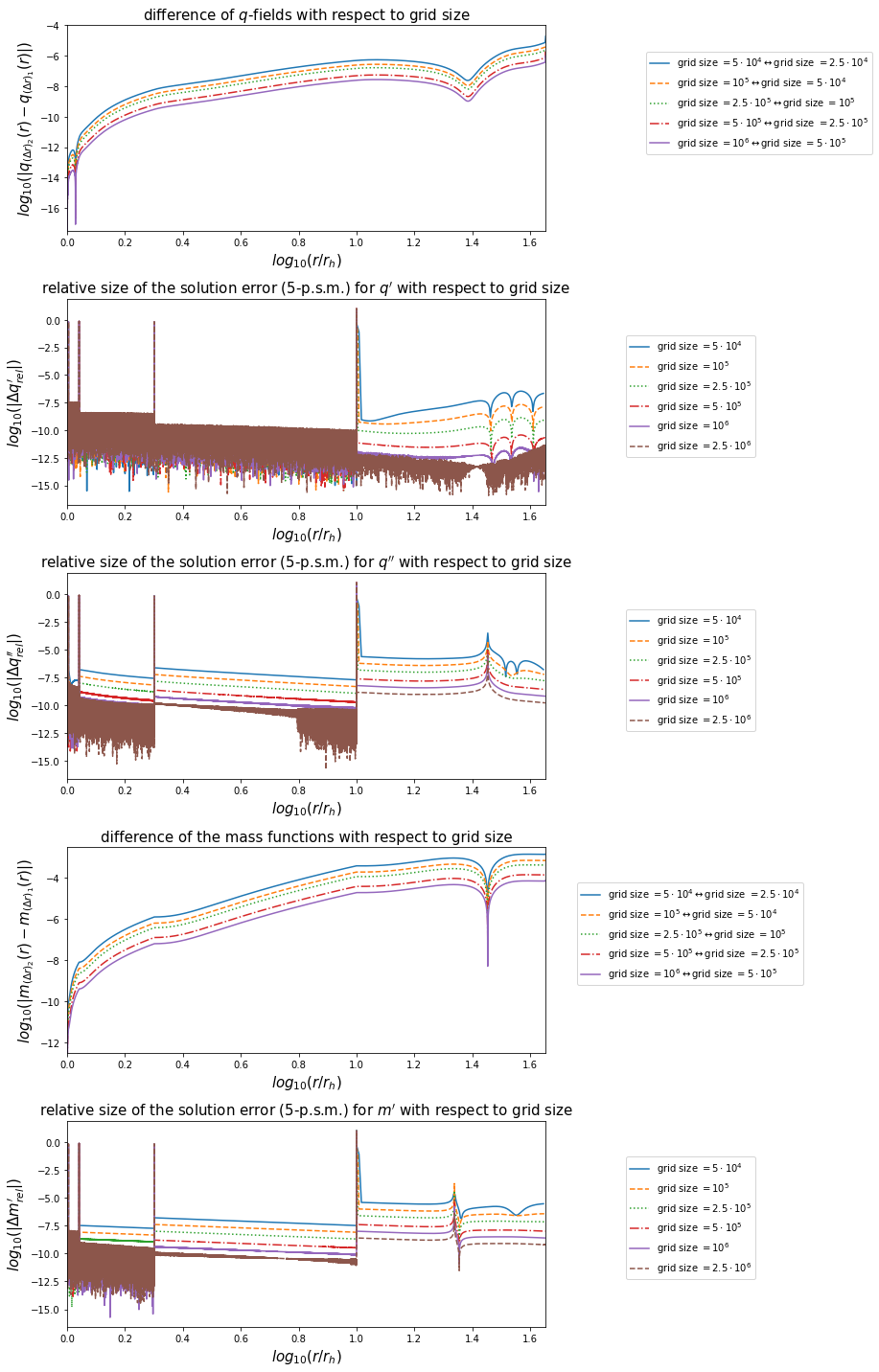}
	\end{center}
	\caption{The effect of different grid sizes on SHBH solutions for $q$-theory outside the event horizon are shown for a horizon radius of $r_h=1$, $q$-field potential parameters $\lambda =1$ and $q_{eq}=0.158$ and shift parameter $\epsilon =10^{-6}$.}
	\label{ppssr6}
\end{figure}
The numerical solution presented in Fig. (\ref{ppssr7}) is approximate both because of finite grid size and finite shift parameter. The solution with $r_h=1$ and $q$-field potential parameters $\lambda =1$ and $q_{eq}=0.158$ is analyzed in Fig. (\ref{ppssr5}) and Fig. (\ref{ppssr6}) with respect to variations of the shift parameter and grid size, respectively.\\
The plots in (\ref{ppssr5}) show that absolute differences of $q$-fields and mass functions for different neighboring shift parameters $\epsilon$ are almost exactly coincident with the absolute size of the imaginary part of the $q$-fields and mass functions for the larger shift parameter present in the corresponding absolute difference plots. The maximal value of the absolute size of the imaginary parts of the represented $q$-fields and mass functions shrinks by one order of magnitude for each decrease of the shift parameter by one order of magnitude as expected. It is about two order of magnitude smaller than the shift parameter for the $q$-fields and about one order of magnitude larger than that of the shift parameter for the mass functions, though. We choose a shift parameter of $\epsilon =10^{-6}$ in most of our plots, as it is seen to be negligibly small to have any effect. The same choice argument will be applied for the grid size to which we now turn.\\
The plots in (\ref{ppssr6}) show the absolute differences of $q$-fields and mass functions for different neighboring grid sizes as well as relative error estimates for the functions $q^{\prime}$, $q^{\prime\prime}$ and $m^{\prime}$ analogous to the lowermost left plot in Fig. (\ref{ppssr7}) for different grid sizes. The expectation that the differences between the $q$-fields and mass functions as well as the error estimates decrease with increasing grid size are fulfilled. The differences of the $q$-fields and mass function decrease by about one order of magnitude for a grid size increase of one order of magnitude. The error estimates are comparable for the different grid sizes close to the horizon. This indicates that no mayor improvement may be achieved with further increase of the grid size. Further away from the horizon the error estimates indeed decrease visibly with increasing grid size.\par 

\section*{Appendix C. Self-consistency.}

\begin{figure}
	\begin{center}
		\includegraphics[scale=0.3]{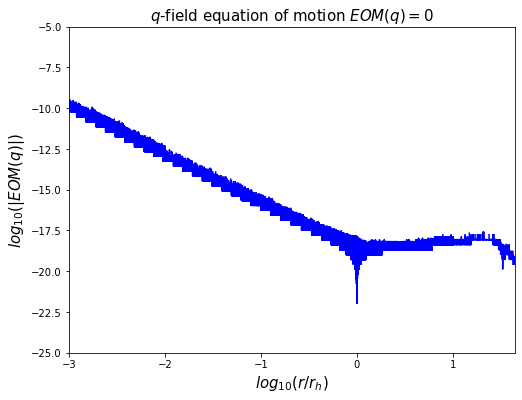}
	\end{center}
	\caption{The self-consistency condition $EOM(q)=0$ for the representative SHBH solution for $q$-theory both outside and inside the event horizon is shown for a horizon radius of $r_h=1$, $q$-field potential parameters $\lambda =1$ and $q_{eq}=0.158$, a grid size of $10^6$ and shift parameter $\epsilon =10^{-6}$.}
	\label{qfieldeomplot}
\end{figure}

We show here the self-consistency of our numerical solutions for the metric functions and the $q$-field. In the process of setting up the field equations we combined the metric field equations with the $q$-field equations into one equation, namely Eq. (\ref{modeinstein}). Aside from this combined equation the individual field equations should be satisfied by the numerical solutions as well. We consider in this regard the fulfillment of the $q$-field equation as presented in Eq. (\ref{qfieldeom}). It may be identically reobtained from the conservation of the energy momentum tensor for our solution

\begin{align}
\nabla_{\alpha}T_q^{\alpha\beta}=&-(\frac{d\epsilon (q)}{dq}-\mu_{eq}-\Box q)\nabla^{\beta}q \\
&-(\frac{d\epsilon (q)}{dq}-\mu_{eq}-h^{\prime}q^{\prime}-h(q^{\prime}(\frac{2}{r}+4\pi r(q^{\prime})^2)+q^{\prime\prime}))\nabla^{\beta}q,
\end{align}

as $q^{\prime}\neq 0$. The $q$-field equation of motion is shown in Fig. (\ref{qfieldeomplot}) with the shorthand notation

\begin{align}
EOM(q)=\frac{d\rho}{dq}-\Box q=\frac{d\epsilon}{dq}-\mu_{eq}-\Box q.
\end{align}

As can be seen, the self-consistency condition or equivalently the fulfillment of the $q$-field equation of motion holds true to numerical accuracy up to the neighborhood of the center where the discrepancy increases beyond numerical precision the further the central region is approached. The latter observation has no effect on any of the conclusions drawn within this work, though.


\end{document}